\documentclass[A4paper,fleqn,usenatbib]{mnras}
\usepackage{ae,aecompl}
\usepackage{graphicx}	
\usepackage{amsmath}	
\usepackage{amssymb}	
\usepackage{amsbsy}
\usepackage{float}
\usepackage{color}
\usepackage{txfonts}
\usepackage{times}

\usepackage{morefloats}
\usepackage{multirow}

\usepackage{mathrsfs,bm}
\newcommand{\bcdot}{\ensuremath{%
  \mathchoice%
   {\mskip\thinmuskip\lower0.2ex\hbox{\scalebox{1.5}{$\cdot$}}\mskip\thinmuskip}}%
   {\mskip\thinmuskip\lower0.2ex\hbox{\scalebox{1.5}{$\cdot$}}\mskip\thinmuskip}%
   {\lower0.3ex\hbox{\scalebox{1.2}{$\cdot$}}}%
   {\lower0.3ex\hbox{\scalebox{1.2}{$\cdot$}}}%
}

\newcommand{\Msun}{\>{\rm M_\odot}}

\newcommand{\kpch}{\rm\>h^{-1}{\rm {kpc}}}
\newcommand{\kpc}{{\rm \ {kpc}}}

\newcommand{\magarcsec}{\rm \ mag\ arcsec^{-2}}

\title[METALLICITY OF LSBGS IN SIMULATIONS]{Stellar Metallicity of Galaxies: New Insight on the Formation and Evolution of Low Surface Brightness Galaxies in the IllustrisTNG Simulation}
\author[L. Tang]{Lin Tang$^{1,2}$\thanks{tanglin23@cwnu.edu.cn}
  \vspace*{0.2cm}  \\
  $^1$School of Physics and Astronomy, China West Normal University, ShiDa Road 1, 637002, Nanchong, China
   \vspace*{0.2cm}  \\
   $^2$CSST Science Center for the Guangdong-Hongkong-Macau Greater Bay Area, DaXue Road 2, 519082, Zhuhai, China
  }
\date{Accepted 2024 March 21. Received 2024 March 21; in original form 2023 December 28}
\pubyear{2024}
\begin{document}

\label{firstpage}
\pagerange{\pageref{firstpage}--\pageref{lastpage}}

\maketitle

\begin{abstract}
In this work, we investigate the stellar metallicities of low surface brightness galaxies (LSBGs) and normal high surface brightness galaxies (HSBGs) in the IllustrisTNG100-1 simulation. 
LSBGs and HSBGs are classified as galaxies with mean central surface brightness $\mu_{\rm r} > 22.0\magarcsec$ and $\mu_{\rm r} < 22.0\magarcsec$, respectively.
Our findings indicate that both LSBGs and HSBGs exhibit similar number distributions of stellar metallicities at high redshifts ($z>1.5$). 
However, at low redshifts ($z<1.5$), a clear bimodal distribution of stellar metallicities in galaxies emerges, with LSBGs tending to be more metal-poor than HSBGs. 
The lower metallicity of LSBGs compared to HSBGs is mostly attributed to the pronounced gradient in the radial distribution of stellar metallicities.
The bimodality of stellar metallicity is not attributed to colour distinctions but rather to the slower metal enrichment in LSBGs compared to HSBGs.
This suggests that the mechanisms driving metal enrichment in LSBGs differ from those in HSBGs. 

\end{abstract}


\section{Introduction}
\label{introduction}
\cite{Freeman1970} reported that the central surface brightness of his sample exhibited a number distribution peaking at $\mu_{\rm B}=21.65\magarcsec$, known as the Freeman Law.
Subsequent advancements in observational technology have led to the discovery of a substantial number of galaxies with $\mu_B$ fainter than $21.65\magarcsec$ \citep[e.g.,][ and the references therein]{Swaters2003, Kuzio2008, Koda2015, Du2015, Roman2017, Cao2023, Fu2023}, which are commonly defined as low surface brightness galaxies (LSBGs).
These findings indicate that the Freeman Law is significantly influenced by observational limitations and does not accurately fit the actual central surface brightness distribution of galaxies \citep[e.g.,][]{Disney1976,Allen1979}.

LSBGs have been extensively studied and are crucial components in the entire comprehension of galaxy formation and evolution \citep[][for reviews]{Impey1997, Bothun1997}.
They have been discovered across various cosmic environments \citep[e.g.,][]{Koda2015,Merritt2016,Roman2017}, contributing significantly, over 10\%, to the mass at the faint end of the luminosity function \citep[e.g.,][]{Minchin2004, Hayward2005, Haberzettl2007}.
These galaxies exhibit notably flat mass densities in their innermost regions with a core \citep[e.g.,][]{Swaters2003, Kuzio2008}.

Compared to the normal ``classical'' galaxies with high surface brightness, known as high surface brightness galaxies (HSBGs), LSBGs display larger radii, higher gas fractions, lower star-forming rates, lower masses, and higher spins \citep[e.g.,][]{Kim&Lee2013, Huang2014, Du2015, Enrique&Bernardo2019, Salinas&Galaz2021, Luis2022}.
Initially discovered in the blue colour sequence \citep[e.g.,][]{deBlok1995, Heller2001, Adami2009}, subsequent studies have revealed some red LSBGs \citep[e.g.,][]{O'Neil1997, Haberzettl2007, Zhong2008, Fu2023}.
LSBG morphology encompasses all types in the Hubble sequence \citep[e.g.,][]{McGaugh1995b, deBlok1999, Auld2006, Shao2015}.
They are generally regarded as more metal-poor than HSBGs, exhibiting a broad distribution ranging from $1/5 \ Z_{\odot}$ to $Z_{\odot}$ \citep[e.g.,][]{McGaugh1994, Galaz2006, Liang2010, Junais2023}.

Multiple investigations have highlighted that the formation mechanisms and evolutionary pathways of LSBGs are diverse from HSBGs.
For instance, \cite{Dalcanton1997} suggested a gravitationally self-consistent model for disc collapse as a function of the mass and angular momentum of the initial protogalaxy. 
In their model, the intrinsic property of low surface brightness in LSBGs is naturally formed in the low-mass and/or high angular momentum halos.
\cite{Jimenez1998} modelled LSBs as non-self-gravitating discs inside isothermal dark matter halos, and confirmed that the formation model with a high spin parameter can explain the colours, the colour gradients, and the chemical abundances of LSBGs.
\cite{vanDokkum2015} suggested that LSBGs might originate from gas stripping from Milky Way-like galaxies.
The differing evolutionary trajectories of LSBGs and HSBGs are also thought to be attributed to the host dark matter halo in which the galaxies reside \citep[e.g.,][]{Roman2017}.
\cite{McGaugh2021} proposed two hypotheses for the galaxy evolution: dependence on the spin or density, and emphasised on the importance of auxiliary hypotheses like feedback or Modified Newtonian Dynamics (MOND) for the formation of LSBGs.
Commonly proposed formation mechanisms for LSBGs involve their emergence in low-mass dark matter halos, driven by physical processes such as unusually rapid rotation, gas ejections triggered by supernova feedback, and tidal stripping during merger events \citep[e.g.,][]{Ruiz-Lara2018, Martin2019, Wright2021}.

Several numerical simulations, including the NIHAO project \citep{Wang2015}, Horizon-AGN \citep{Dubois2014}, EAGLE project \citep{Schaye2015}, IllustrisTNG project \citep{Springel2018}, have effectively simulated certain types of LSBGs \citep[e.g.,][]{DiCintio2019, Martin2019, Kulier2020, Luis2022}. 
These studies have theoretically reproduced the observational characteristics of LSBGs and provided various physical mechanisms for their formation. 
For instance, it is found that the dark matter halo hosting LSBGs exhibits high-speed rotation, thereby contributing to the extended disc structure of LSBGs \citep[e.g.,][]{Martin2019, Kulier2020, Luis2022}.
Processes such as mergers and gas accretion likely play significant roles in the formation of LSBGs \citep[e.g.,][]{DiCintio2019}.

In the NIHAO project, \cite{DiCintio2017} explored the origin of ultra-diffuse galaxies (UDGs), which correspond to the dwarf population of LSBGs characterized by a mean central surface brightness fainter than $24\magarcsec$.
They highlighted that the presence of gas is essential for the formation of UDGs.
\cite{Sales2020} and \cite{Benavides2021} investigated the formation of UDGs in the IllustrisTNG simulation, and found that the interactions and tidal stripping lead to the transition of UDGs from star forming to isolated quiescent.
\cite{Zhu2018} identified a Malin 1-like giant LSBG in the IllustrisTNG simulation, exhibiting good agreement with observations.
Subsequently, \cite{Zhu2023} reported a giant LSBG sample of around 200 galaxies in the IllustrisTNG simulation.
They determined that the aligned mergers and isolated environments are crucial factors in the formation of extended discs. 

In this work, we focus on the stellar metallicities of LSBGs and HSBGs, their correlations with galaxy properties, and redshift evolution of the stellar metallicities, aiming to offer an alternative insight into the formation of LSBGs.
In Sections ~\ref{Simulation and Galaxy Sample}, we provide a briefly description of the simulation and the galaxy samples used.
Our primary findings are presented in Section ~\ref{Results}.
Finally, we provide a brief summary of our conclusions, in Section \ref{Conclusions}.
\begin{figure*}
    \centering
\includegraphics[width = 0.3\textwidth]{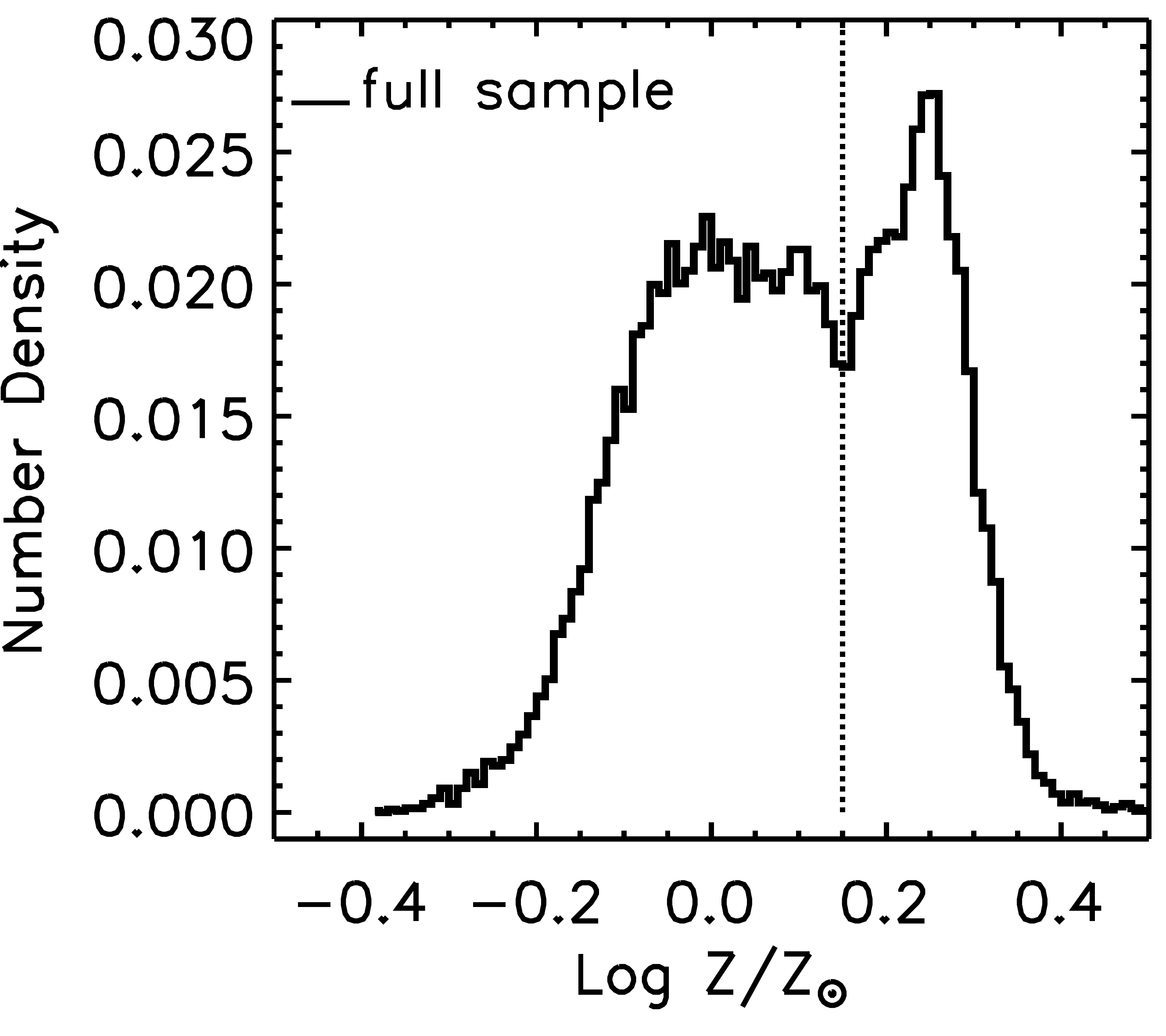}
\includegraphics[width = 0.3\textwidth]{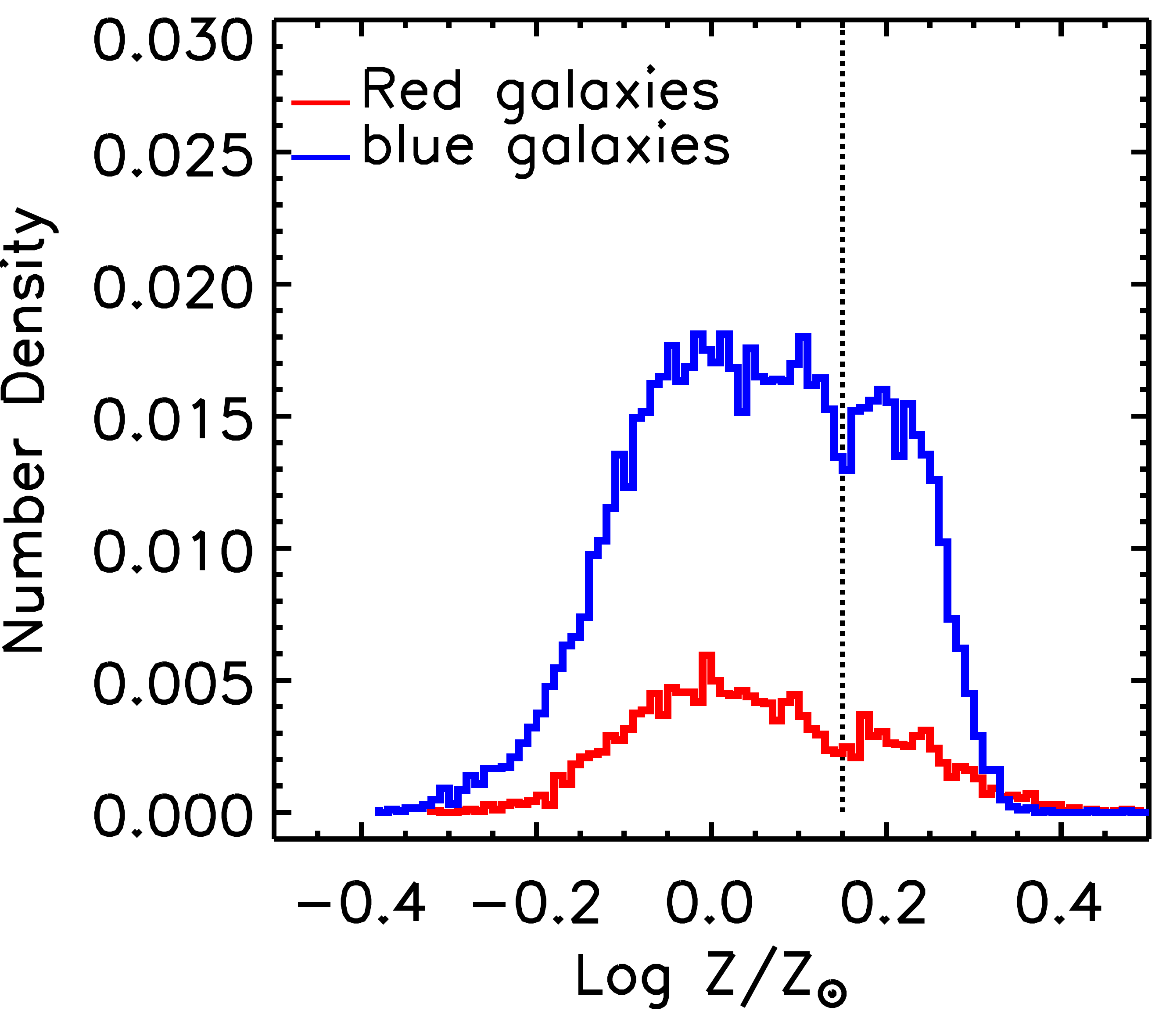}
\includegraphics[width = 0.3\textwidth]{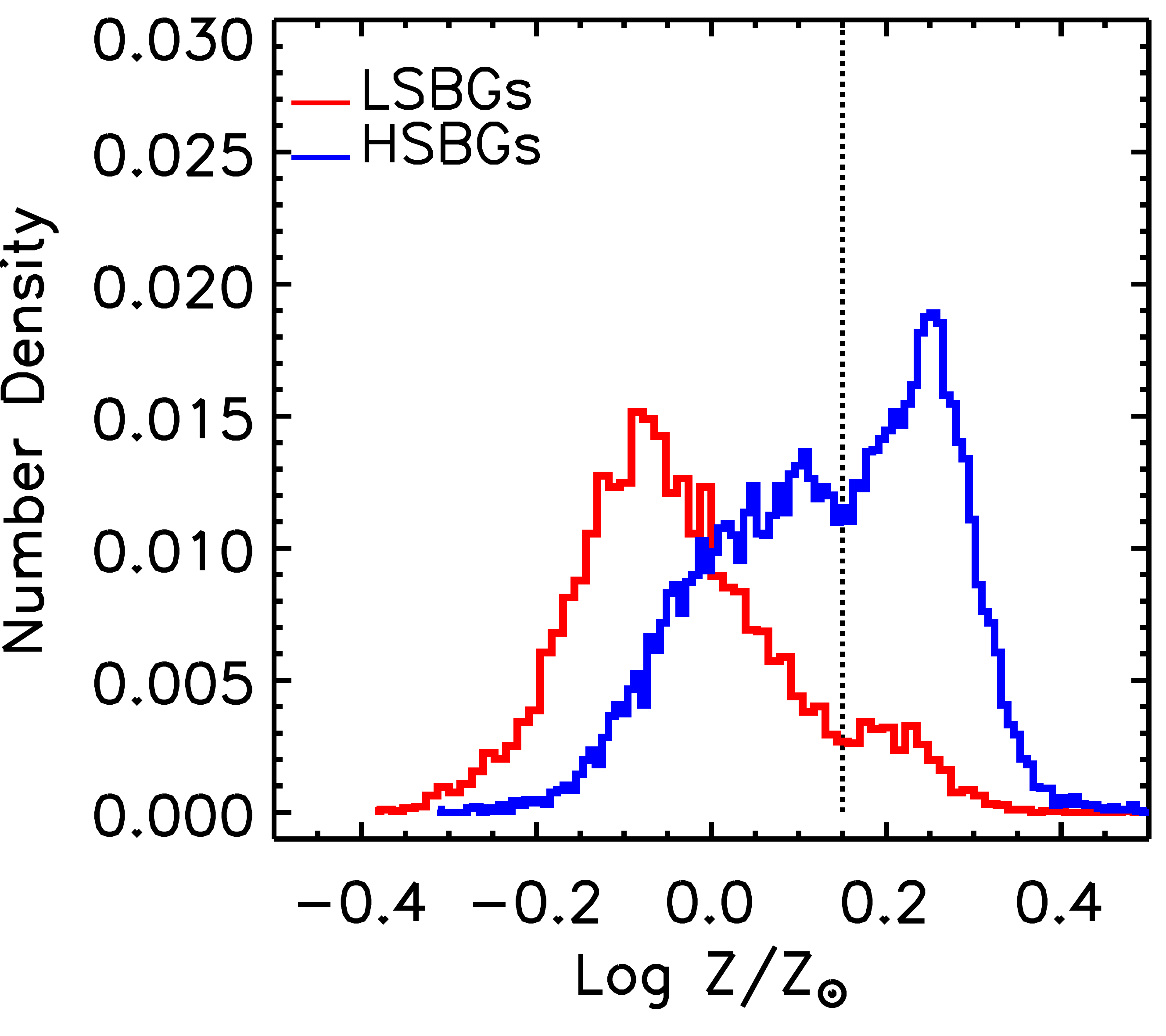}
\caption{
Number density of the stellar metallicity for the full sample (left panel), red and blue galaxies (middle panel), LSBGs and HSBGs (right panel).
The colour lines represent the subsamples as shown in the labels of each panels. 
The dotted lines denote $\log Z/Z_{\odot}=0.15$.
Note that the bin sizes in those histograms are not same, set as the $autobin$.
}
  \label{Metallicity_distribution}
\end{figure*}
\begin{figure}
    \centering
\includegraphics[width = 0.3\textwidth]{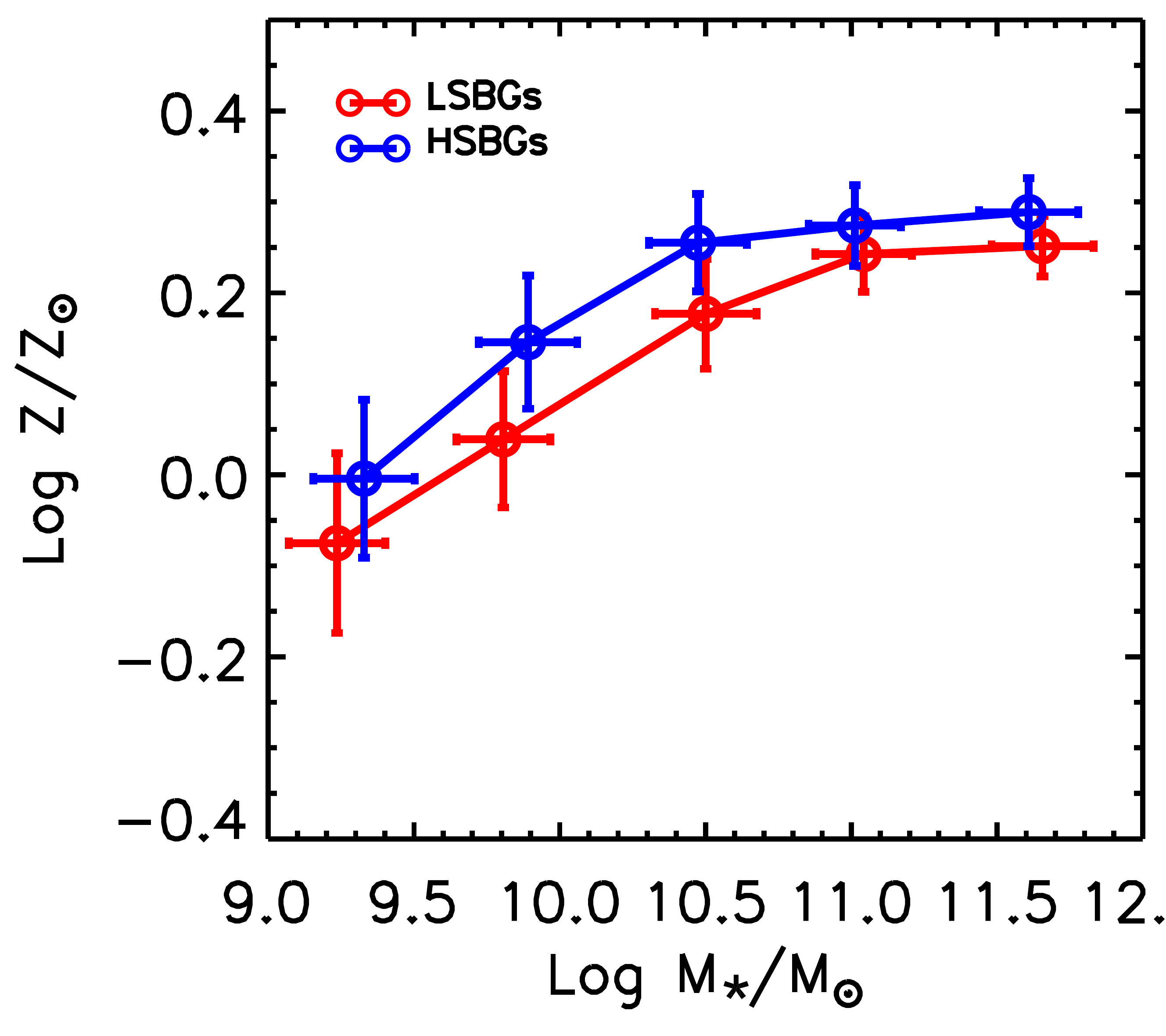}
\caption{Stellar Mass--Metallicity relationships: red plots for LSBGs and blue plots for HSBGs. 
We divide the sample into five subsamples within the stellar mass regions, $9.0<\log M_{\star}/\Msun<9.6$, $9.6<\log M_{\star}/\Msun<10.2$, $10.2<\log M_{\star}/\Msun<10.8$, $10.8<\log M_{\star}/\Msun<11.4$, $11.4<\log M_{\star}/\Msun<12.0$.
}
  \label{Metal_Mass}
\end{figure}
\begin{figure}
    \centering
\includegraphics[width = 0.3\textwidth]{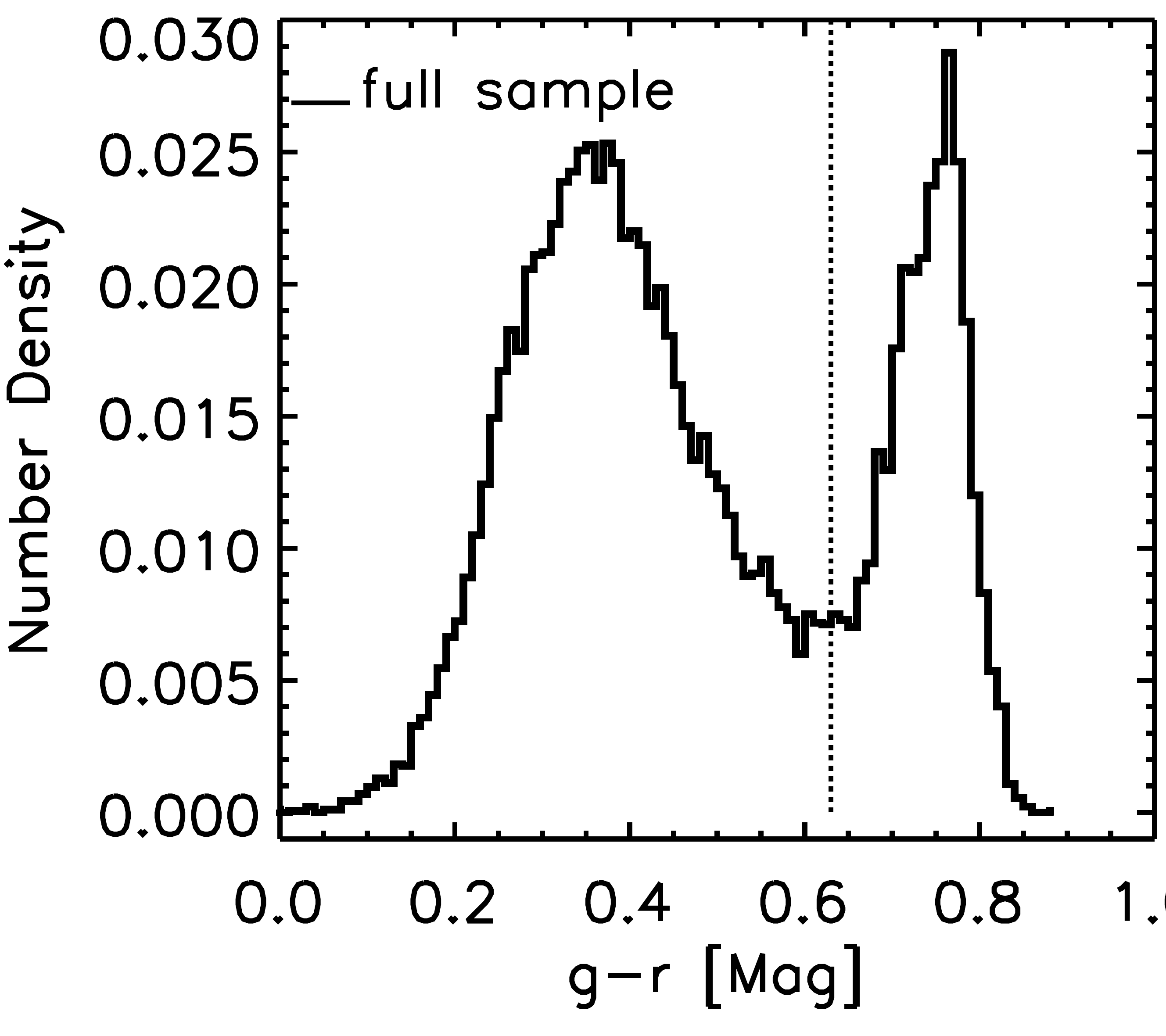}
\caption{
Number density of the g-r colour of the full sample. 
The dotted line denotes $g-r = 0.63$.
}
  \label{Color_distribution}
\end{figure}
\begin{figure*}
    \centering
\includegraphics[width = 0.75\textwidth]{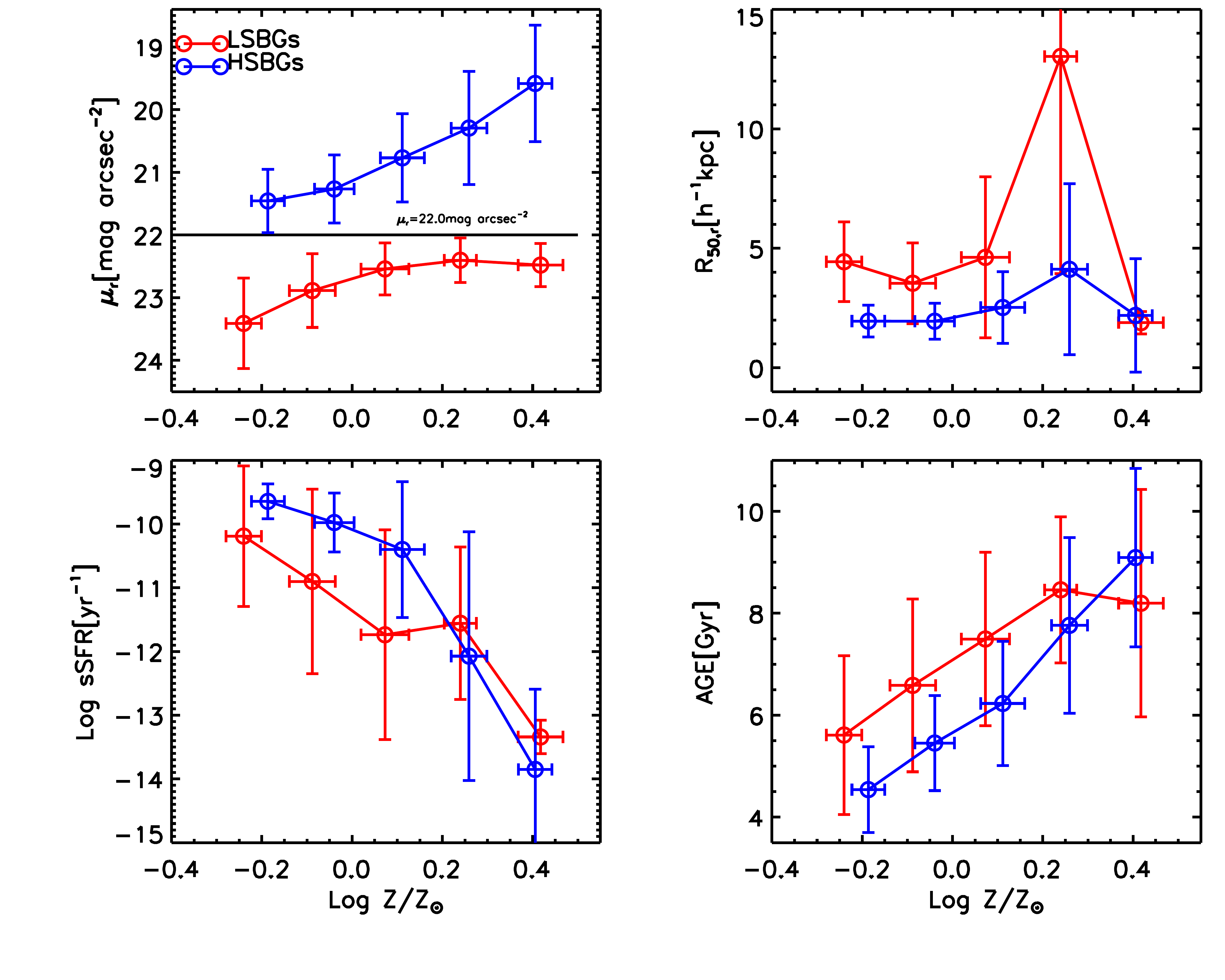}
\caption{
Relationships between the stellar metallicities of LSBGs (red lines with circle symbols) and HSBGs (blue lines with circle symbols) and the mean central surface brightness brightness $\mu_r$, half light radius $R_{\rm 50,r}$, specific star formation rate $sSFR$, and mean age of all the stellar particles $AGE$. 
The solid black line in the top-left panel is the value we discriminate LSBGs from HSBGs, i.e., $\mu_r=22.0\magarcsec$.
The error bars are the standard deviation. 
}
  \label{Metallicity_properties}
\end{figure*}
\begin{figure}
    \centering
\includegraphics[width = 0.35\textwidth]{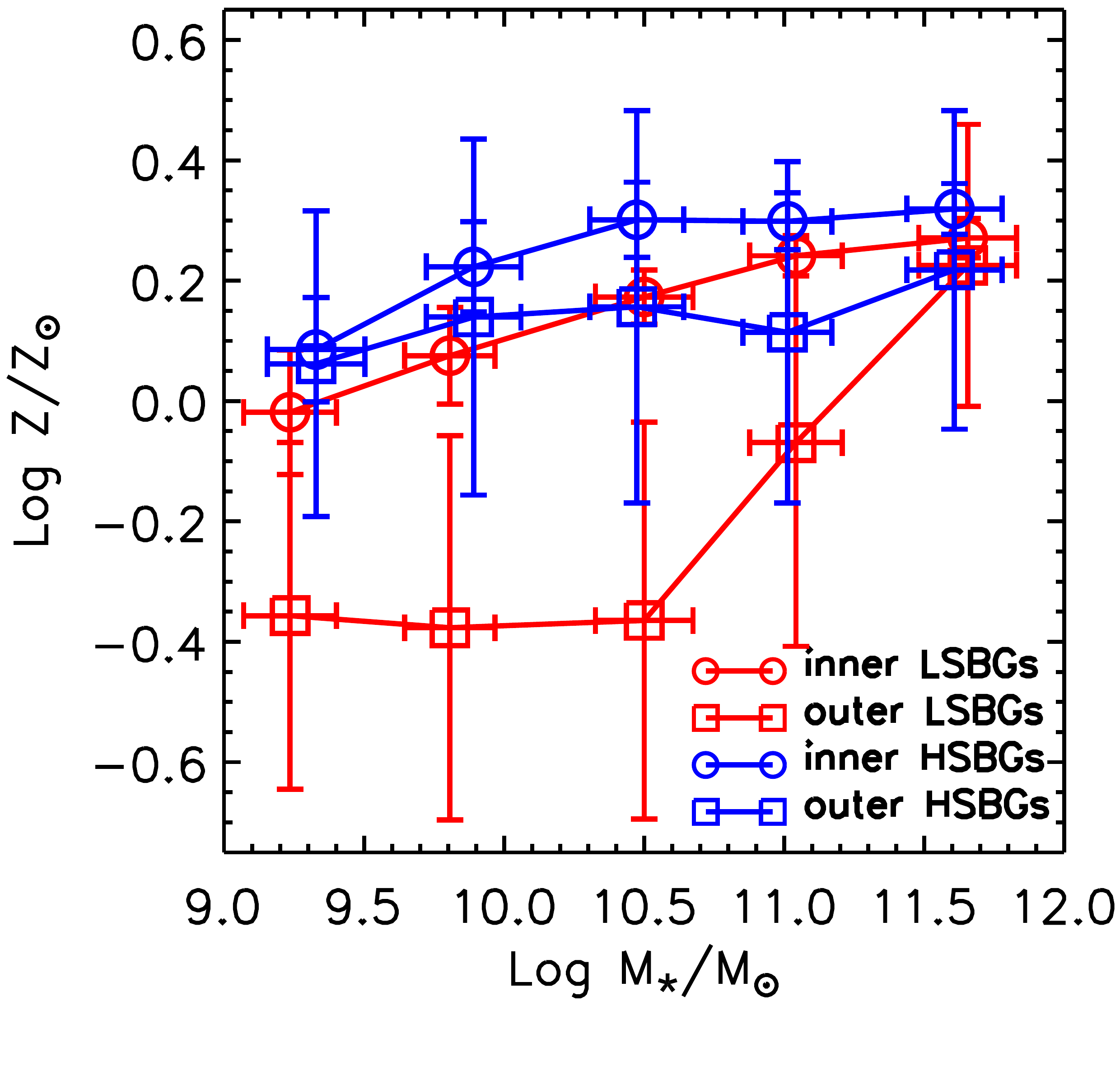}
\caption{
Relationships between the stellar metallicity ($\log Z/Z_{\odot}$) and stellar mass ($\log M_{\star}/\Msun$) of galaxies.
Red lines are the results of LSBGs, while blue lines are for HSBGs.
Circle symbols represent the $\log Z/Z_{\odot}$--$\log M_{\star}/\Msun$ relationship of inner region ($R<R_{\rm 50,r}$).
Square symbols are plotted for the outer region ($R>R_{\rm 50,r}$).
}
  \label{radial_distribution_inandout}
\end{figure}
\begin{figure*}
    \centering
    \includegraphics[width = 0.9\textwidth]{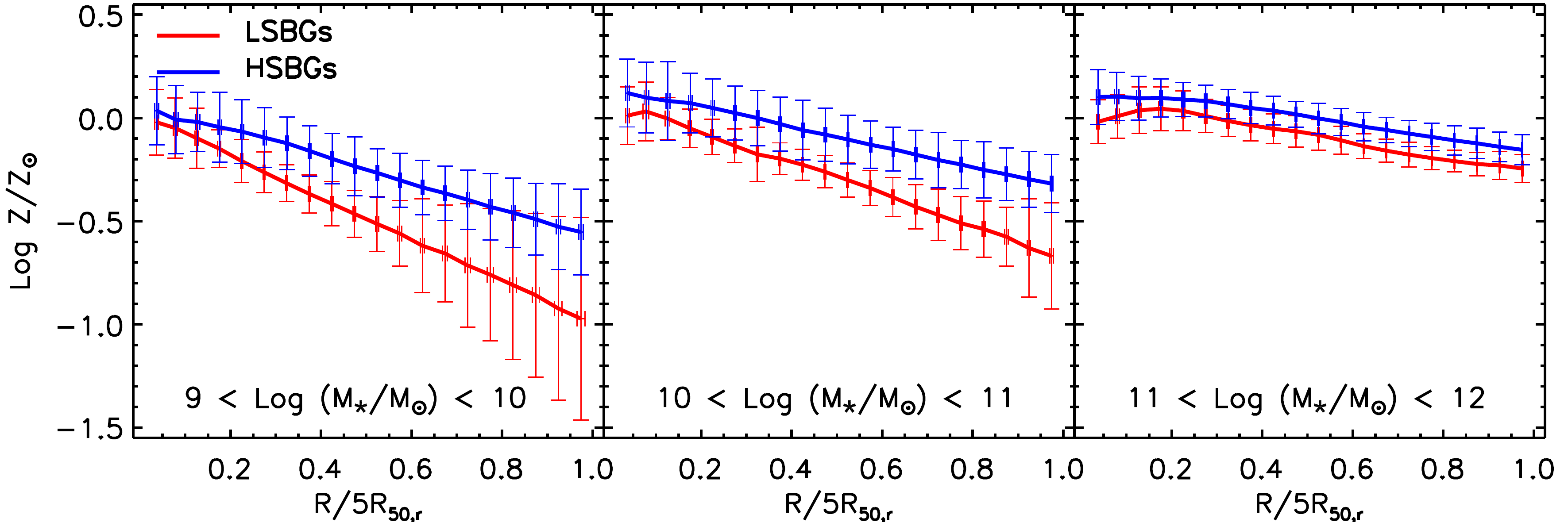}
\caption{
Radial distributions of the stellar metallicity of LSBGs (red lines) and HSBGs (blue lines).
We divide the results into three stellar mass bins of $9<\log M_{\star}/\Msun < 10$, $10<\log M_{\star}/\Msun < 11$, $11<\log M_{\star}/\Msun < 12$, from left to right panel.
}
  \label{radial_distribution_profile}
\end{figure*}
\section{Simulation and Galaxy Sample}\label{Simulation and Galaxy Sample}
In this work, the simulation used is the TNG100-1 simulation, i.e., the highest resolution version of the TNG100 simulation of IllustrisTNG suite \footnote{https://www.tng-project.org}.
The detailed descriptions of the database can be found in \cite{Pillepich2018, Springel2018, Nelson2018, Naiman2018, Marinacci2018, Nelson2019}. 
The TNG100-1 simulation includes a revised active galactic nucleus (AGN) feedback model to control the star formation efficiency of massive galaxies \citep{Weinberger2017}, and a galactic wind mode to inhibit the efficient star formation in low- and intermediate-mass galaxies \citep{Pillepich2018}, compared to the previous Illustris simulations \citep[][]{Vogelsberger2014}.
The dark matter halos are identified by a Friends-of-Friends algorithm \citep[FoF,][]{Davis1985}.
The subhalos are identified as overdense, gravitationally bound substructures with the SUBFIND algorithm \citep{Springel2001, Dolag2009}.
The simulation has a mass resolution for baryon particles of $1.4\times10^6\Msun$ and a Plummer softening length of $0.74\kpc$.

To mitigate the resolution effects, the galaxies selected are the subhalos with stellar masses $M_{\star} > 10^9 \ \Msun$ and half light radius $R_{\rm 50,r} > 1\ \kpch$ in the r-band of the Sloan Digital Sky Survey (SDSS), utillizing $h=0.6774$.
$R_{\rm 50,r}$ refers to the comoving radius of a sphere centered at the comoving center of mass of a subhalo, while this sphere is defined such that half of the total light in the SDSS-r band is included within it.
The mean central surface brightness $\mu_{\rm r} $ is computed using Equation 1 in \cite{Luis2022}.
The view, redshift, fluxes, and dust attenuation of the galaxies are treated as same as those in \cite{Luis2022}.
Additionally, galaxies with $\log M_{\star}/\Msun > 12$ are excluded in the samples, as they are very extended components, making their classification as LSBGs difficult, primarily due to the strong influence and inclusion of the intracluster light (ICL) component \citep{Luis2022}.

The final full sample consists of 18,674 galaxies at snapshot 99 ($z=0.0$), including 5,334 LSBGs and 13,340 HSBGs, categorized by a threshold mean central surface brightness of $\mu_{\rm r} = 22.0\magarcsec$ \citep{Luis2022}.
Note that the number of galaxies in the full sample is smaller than that in \cite{Luis2022}, {due to} the additional criterion involving the half light radius $R_{\rm 50,r}$.
These substructures with $R_{\rm 50,r} < \ 1\kpch$ have lower mass with $<\log M_{\star}/\Msun > \sim 9.3$, but are roughly four magnitudes brighter in the central region than their counterparts with larger radii.
The $\mu_r$--$M_{\star}$ relationship of the substructures with $R_{\rm 50,r} < 1\kpch$ deviates from that of the full sample.
Furthermore, a majority of these substructures lack identifiable progenitors at high redshifts.

\section{Results}\label{Results}
\subsection{Number Density}\label{number distribution}
In the IlustrisTNG-1 simulation dataset, the stellar metallicity is computed as the mass-weighted average metallicity $M_{\rm Z}/M_{\rm tot}$ of the star particles within twice the half stellar mass radius, where $Z$ is any element above helium.
In the left of Figure ~\ref{Metallicity_distribution}, we present the number density (galaxy numbers in each bin divided by the total number of galaxies) of the stellar metallicity for the full sample.
Note that the bin sizes in these histograms are not same, and set as $autobin$\footnote{We plot the histograms using the IDL routine $plothist$ in \texttt{IDLAstro} with a bin size set of $autobin$, which means that we automatically determines bin size of the histogram as the square root of the number of samples.}.
This distribution notably exhibits double peaks, with a distinct valley at $\log Z/Z_{\odot} \sim 0.15$.
As one of the most important bimodal distribution for the galaxy properties is found in colour \citep[e.g.,][]{Baldry2004, Dekel2006}, we illustrate the stellar metallicity distributions of the red and blue galaxies simply segregated by $g-r=0.63$ in the middle panel of Figure ~\ref{Metallicity_distribution}.
The value of $0.63$ is derived from the number distribution of the galaxy colour as shown in Figure ~\ref{Color_distribution}. 
Remarkably, both red and blue galaxies exhibit a broad range of stellar metallicities, spanning the valley of the bimodal distribution.
This simply implies that the bimodality of stellar metallicity is not attributed to colour distinctions.
In the right panel of Figure ~\ref{Metallicity_distribution}, we present the number density of stellar metallicity for LSBGs and HSBGs.
It is evidently discovered that LSBGs primarily contribute to the peak of low stellar metallicity in the left panel, whereas HSBGs make up of the peak of high one.

LSBGs are generally less massive than HSBGs, and according to the stellar mass--metallicity relation, high-mass galaxies are expected to be metal-rich \citep[e.g.,][]{Delgado2014}.
We explore the stellar mass--metallicity relation of LSBGs and HSBGs, as shown in Figure ~\ref{Metal_Mass}, which indicates that for a given stellar mass, HSBGs are metal richer than LSBGs.
Moreover, we check the histograms of HSBGs and LSBGs for the five sub-samples with different stellar mass regions (not shown in the text).
In each region, the histograms are similar to that in the right panel of Figure \ref{Metallicity_distribution}.
These findings suggest that the bimodality of the stellar metallicity distribution is mainly due to the stellar metallicity distinctions between LSBGs and HSBGs.
The metal-poor characteristic in LSBGs agrees with the previous observations \citep[e.g.,][]{Liang2010, Schombert2021, Junais2023}.

Note that LSBGs and HSBGs cannot be distinguished from the valley in the stellar metallicity number distribution of galaxies.
This goes beyond the straightforward division of the galaxies into red and blue categories.
From the left panel of Figure ~\ref{Metallicity_distribution}, the distribution exhibits a relatively flat region within the range of $-0.10 <\log Z/Z_{\odot} < 0.15$.
This flattened region is notably influenced by the presence of metal-poor HSBGs.
Likewise, there exists a fraction of LSBGs that exhibit a metal-rich composition, resembling the majority of HSBGs.
Consequently, attempting to perfectly classify LSBGs and HSBGs based on a strictly bimodal distribution of the stellar metallicity is not feasible.

\subsection{Relationships between Stellar Metallicity and Galaxy Properties}\label{relationship with galaxy properties}
In this section, we explore the relationships between stellar metallicity and various galaxy properties, as shown in Figure ~\ref{Metallicity_properties}.
These galaxy properties consist of the mean central surface brightness ($\mu_r$), half light radius ($R_{\rm 50,r}$), specific star formation rate ($sSFR=SFR/M_{\star}$), and mean age of all the star particles ($AGE$).
Figure ~\ref{Metallicity_properties} clearly illustrates that both LBSGs and HSBGs exhibit similar relationships between stellar metallicity and these galaxy properties. 
From Figure \ref{Metallicity_properties} and \ref{Metal_Mass}, it is found that more metal-rich galaxies tend to be more massive, lower effective in star formation, and older.
Despite that both LSBGs and HSBGs have similar size-metallicity relations, at fixed stellar metallicity, LSBGs are systematically larger, and also more massive, lower effective in star formation, and older than HSBGs. 
Note that we have set the star formation rate ($SFR$) values to $10^{-4}$ for $SFR=0$.
It is also found that the relationships between stellar metallicity and galaxy properties, especially for $R_{\rm 50,r}$, exhibit significant transitions for the most metal-rich LSBGs.
We carefully check the galaxy sample, and find that there are less than ten LSBGs within the most metal-rich region.
The transitions might be a consequence of statistical deviation.
As shown in the stellar mass--metallicity relation \citep[e.g.,][and Figure 2 in this paper]{Delgado2014}, the metallicity increases with increasing stellar mass.
There is a reasonable doubt that the difference between LSBGs and HSBGs for given stellar metallicity is a consequence of a strong bias caused by the stellar mass, similar to the size and age \citep[e.g.,][]{Shen2003, Behroozi2013}.
However, this is beyond the purpose of this paper and will be investigated in future analyses.

\subsection{Radial Distribution}\label{Radial distribution}
\cite{Kim2012} considered that the metallicity gradient is a valuable tracer of the evolution of LSBGs, yet poorly investigated in observations.
In this section, we present the stellar metallicity of LSBGs at varying radii and compare it with that of HSBGs, as shown in Figure ~\ref{radial_distribution_inandout} and ~\ref{radial_distribution_profile}. 
Figure ~\ref{radial_distribution_inandout} illustrates the mean stellar metallicity within and beyond $R_{\rm 50,r}$.
It is found that within $R_{\rm 50,r}$ denoted by circles symbols, the mean stellar metallicities of LSBGs are slightly lower but closely aligned with those of HSBGs.
However, as indicated by square symbols, the mean stellar metallicities of LSBGs beyond $R_{\rm 50,r}$ are significantly lower than those of HSBGs, particularly for galaxies with intermediate and low stellar mass.
\cite{Luis2022} found that at redshift $z\sim0$, the specific star formation rate ($sSFR$) of LSBGs beyond the effective radius is higher compared to that of HSBGs.
This suggests more efficient star formation in the outer parts of LSBGs, consequently leading to obviously lower stellar metallicity donated by the red square symbols, compared to that of HSBGs donated by the blue square symbols.

The trends in Figure ~\ref{radial_distribution_inandout} are further supported by the radial distributions in Figure ~\ref{radial_distribution_profile}. 
Given the correlation between stellar metallicity and stellar mass in Figure ~\ref{radial_distribution_inandout}, we exhibit the radial distributions for three given stellar mass ranges, arranged from left to right panel of Figure ~\ref{radial_distribution_profile}.
It becomes apparent that the stellar metallicities of LSBGs exhibit a pronounced decrease toward larger radius, with steeper slops compared to HSBGs, especially for galaxies with $\log M_{\star}/\Msun < 11$.
This finding is consistent with observations \citep[e.g.,][]{Young2015, Hemler2021}.
These results suggest that the more metal-poor nature of LSBGs is predominantly caused by a more distinct gradient in the radial distribution of stellar metallicity, compared to HSBGs.

\subsection{Redshift Evolution}\label{redshift evolution}
\begin{table}
\centering
 \begin{tabular}{c c c c}
    \hline
    redshift ($z$) & full sample & LSBGs & HSBGs \\
    \hline
    4&   -0.78& -0.87 & -0.73\\
    3&   -0.53& -0.60 & -0.45\\
    2&   -0.39& -0.45 &-0.34\\
    1.5& -0.33& -0.36 &-0.28\\
    1.0&  --   & -0.25 &  0.25\\
    0.5&  --  & -0.125&  0.25\\
    0.3&  --  & -0.10 &  0.25\\    
    0.2&  --  & -0.09 &  0.25\\
    0.1&  --  &-0.09 &   0.25 \\
    0.0&  --  &-0.08 &   0.25 \\ 
    \hline
        \end{tabular}
\caption{ 
The peaks of the full sample, LSBGs, and HSBGs at ten different redshifts in Figure \ref{Metallicity_distribution_Z} and the right panel of Figure \ref{Metallicity_distribution}. 
In this table, the values for HSBGs at $z \le 1$ are the peaks at metal-rich region.
Note that we only list the peaks for the full sample at $z \ge 1.5$, as the number densities for the full sample at low redshifts ($z\le 1$) show the double peaks.
}
\label{peak_table}
\end{table}
\begin{figure*}
    \centering
\includegraphics[width = 0.3\textwidth]{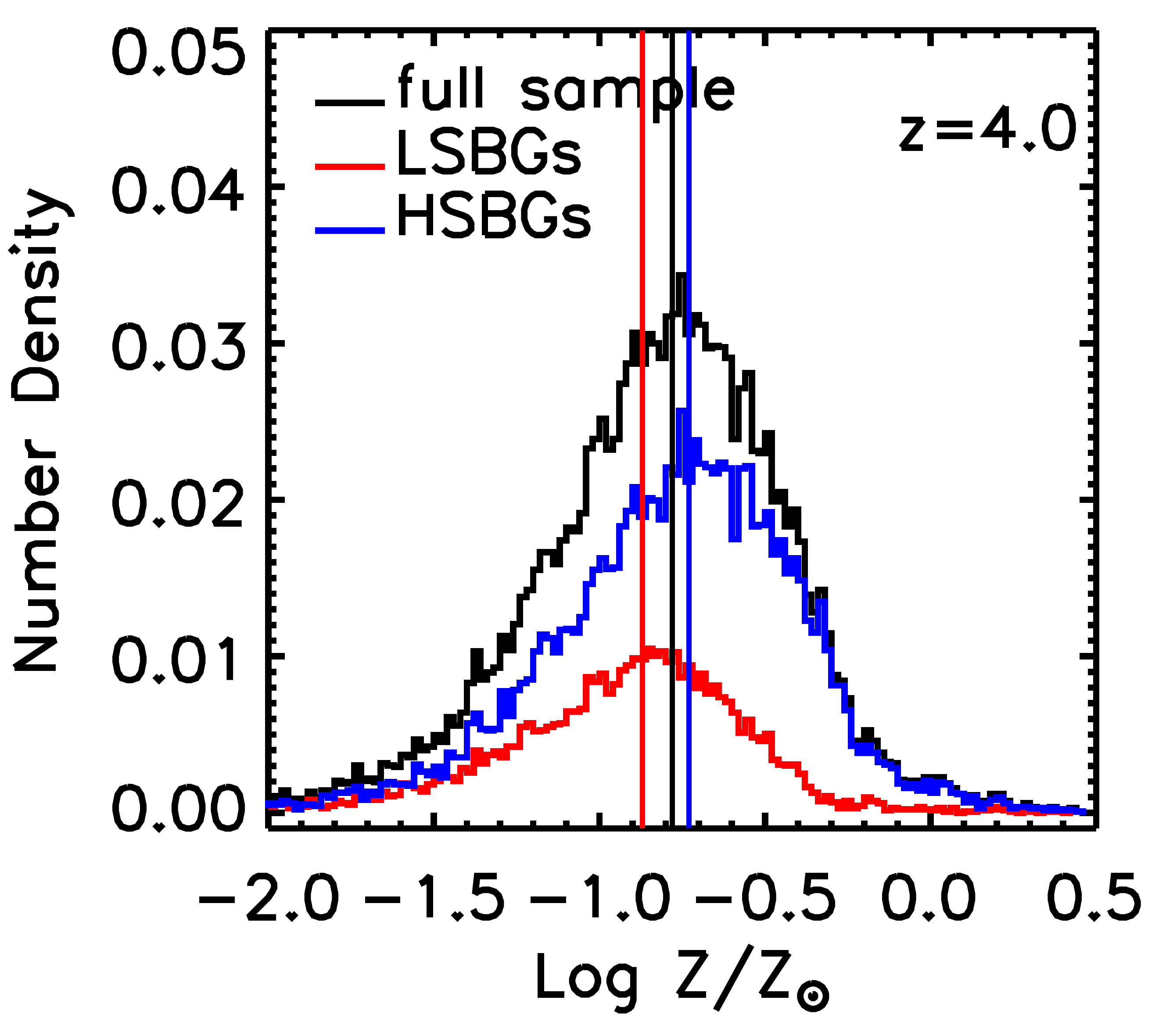}
\includegraphics[width = 0.3\textwidth]{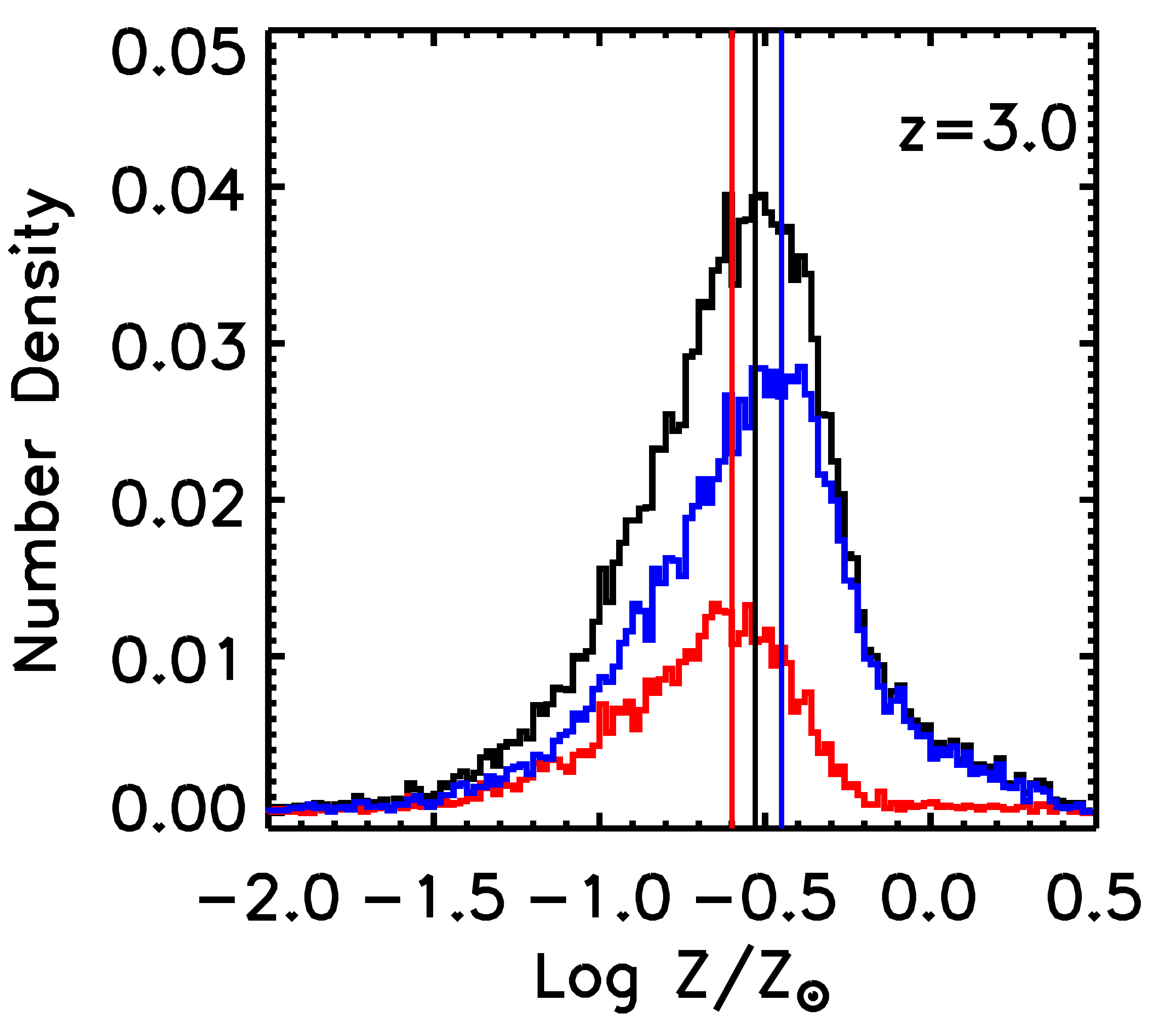}
\includegraphics[width = 0.3\textwidth]{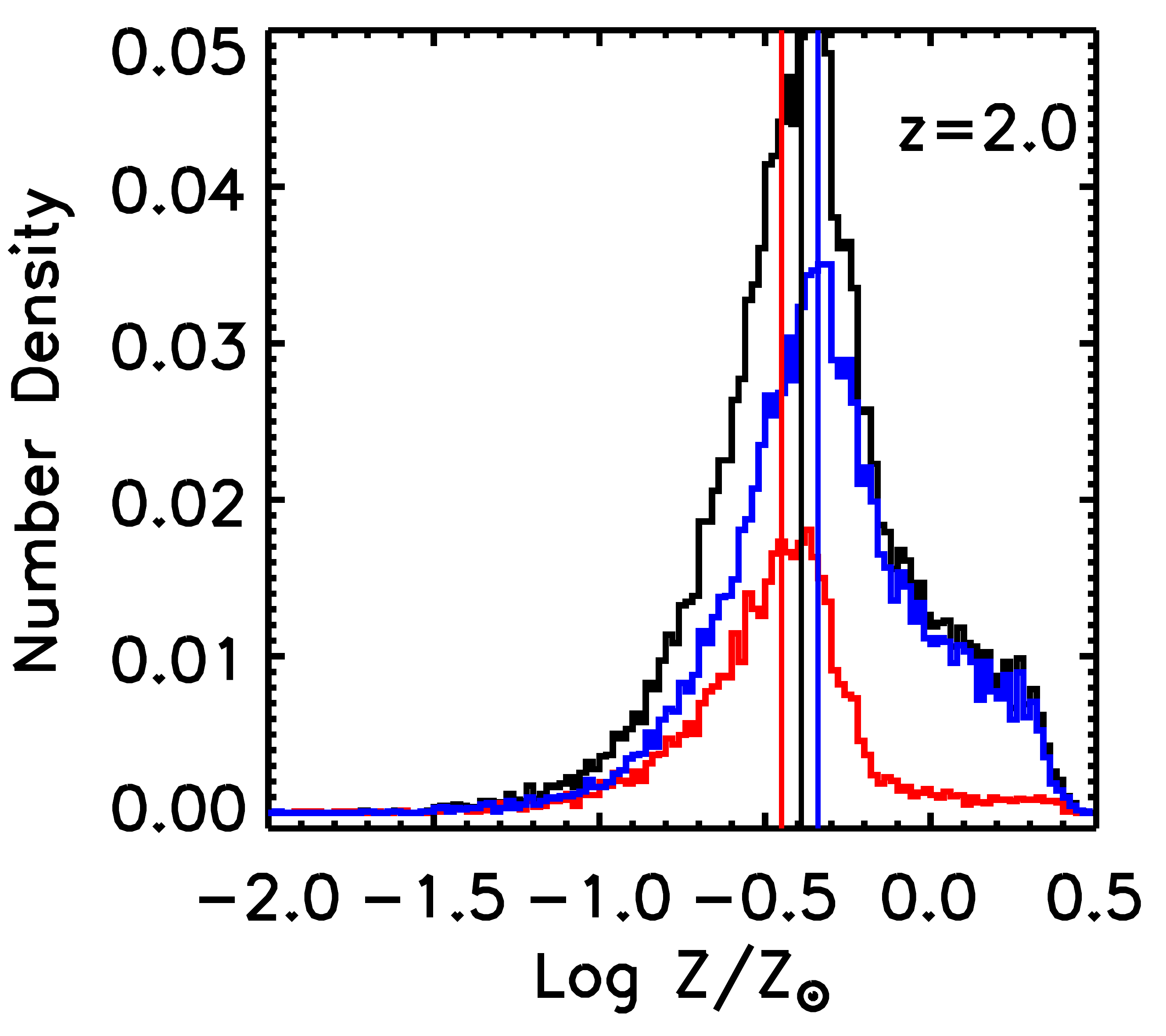}
\includegraphics[width = 0.3\textwidth]{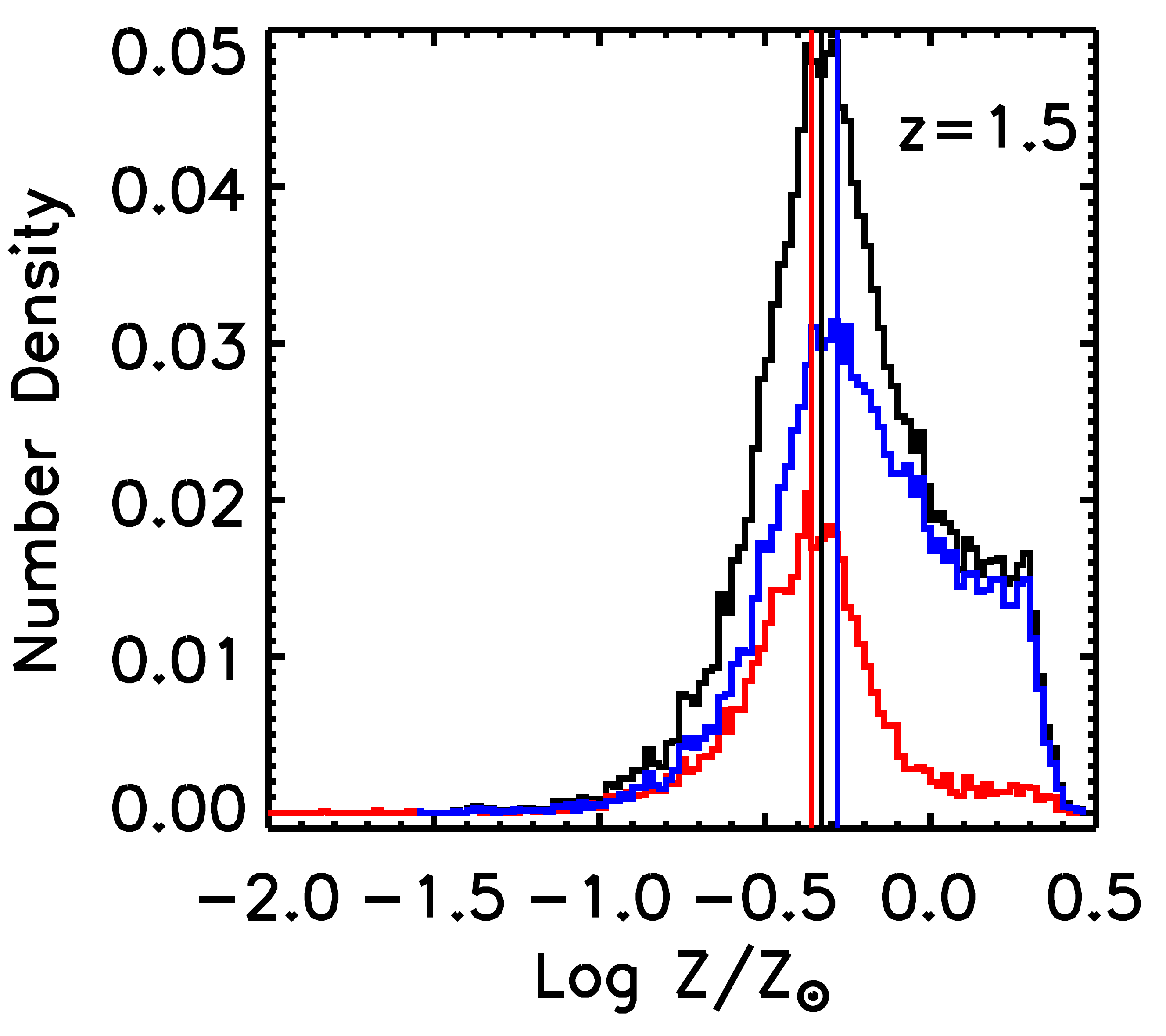}
\includegraphics[width = 0.3\textwidth]{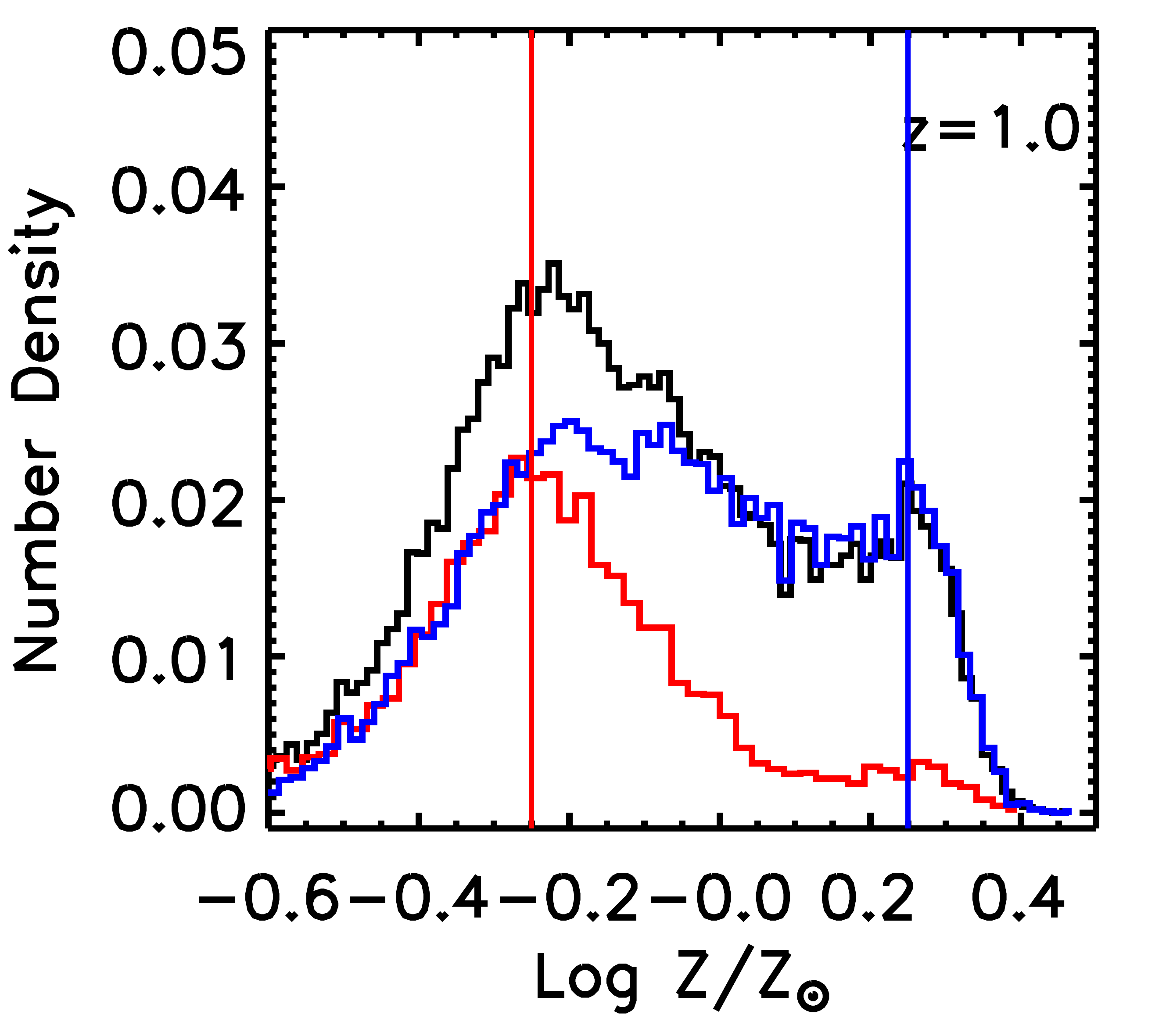}
\includegraphics[width = 0.3\textwidth]{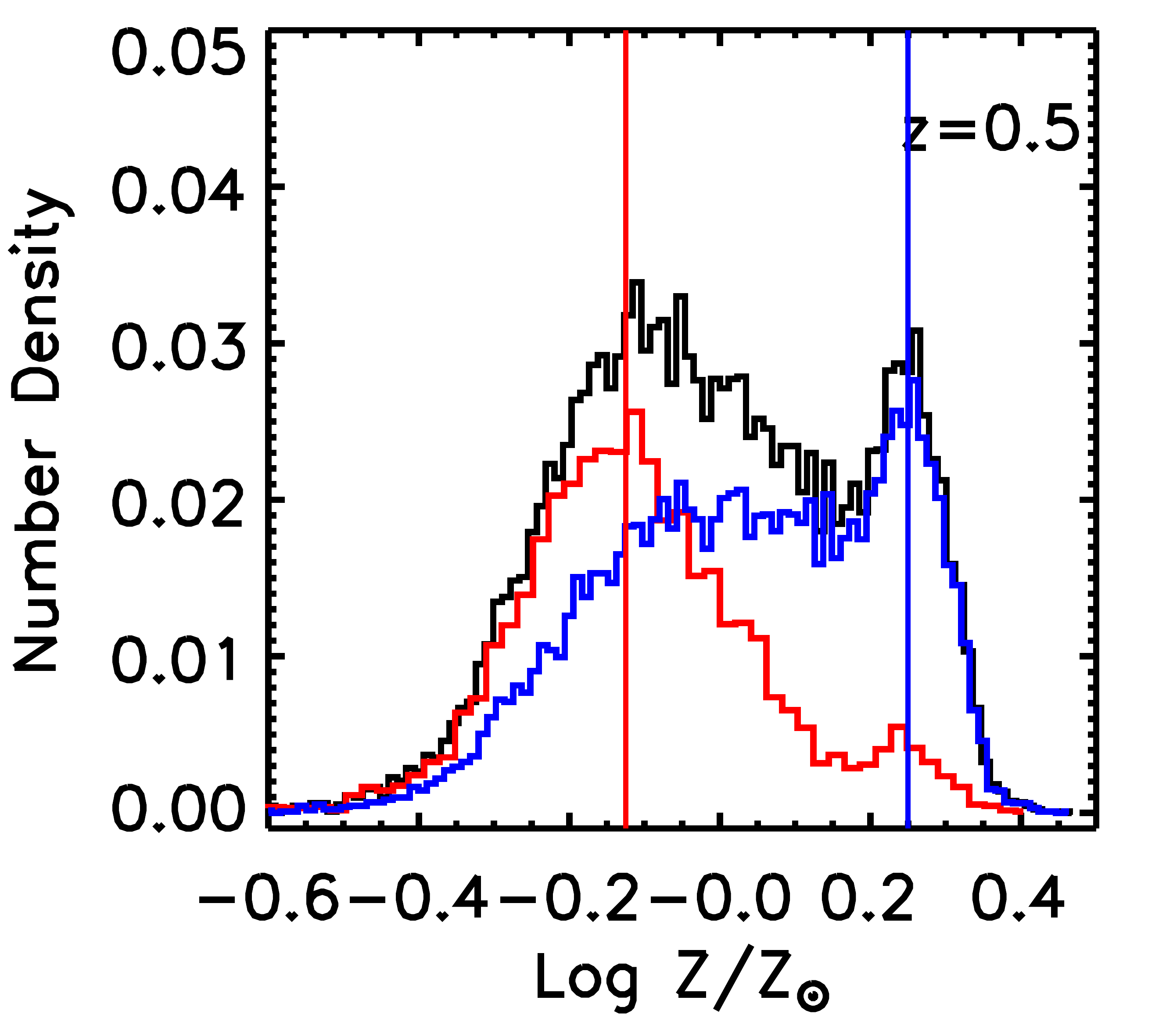}
\includegraphics[width = 0.3\textwidth]{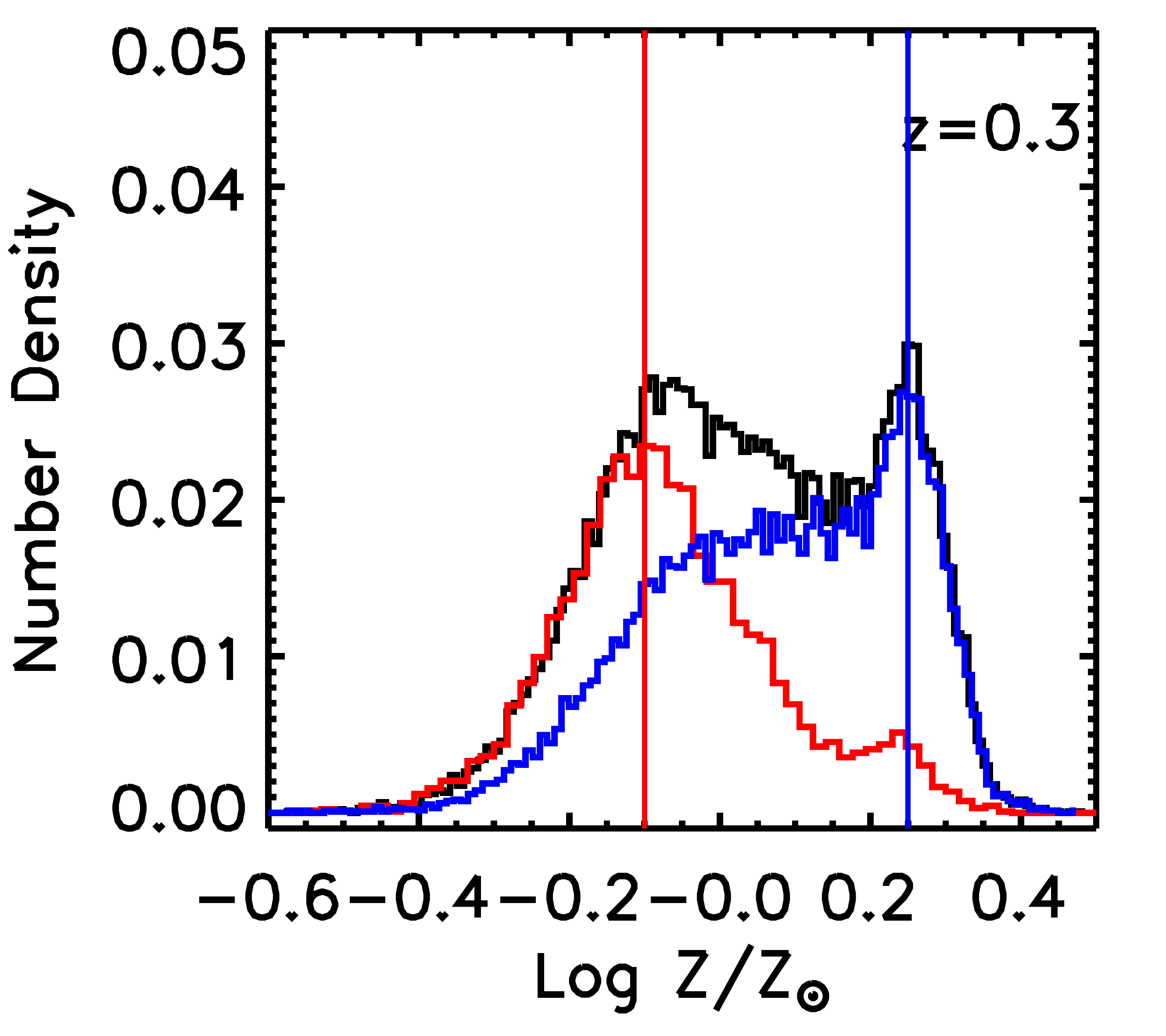}
\includegraphics[width = 0.3\textwidth]{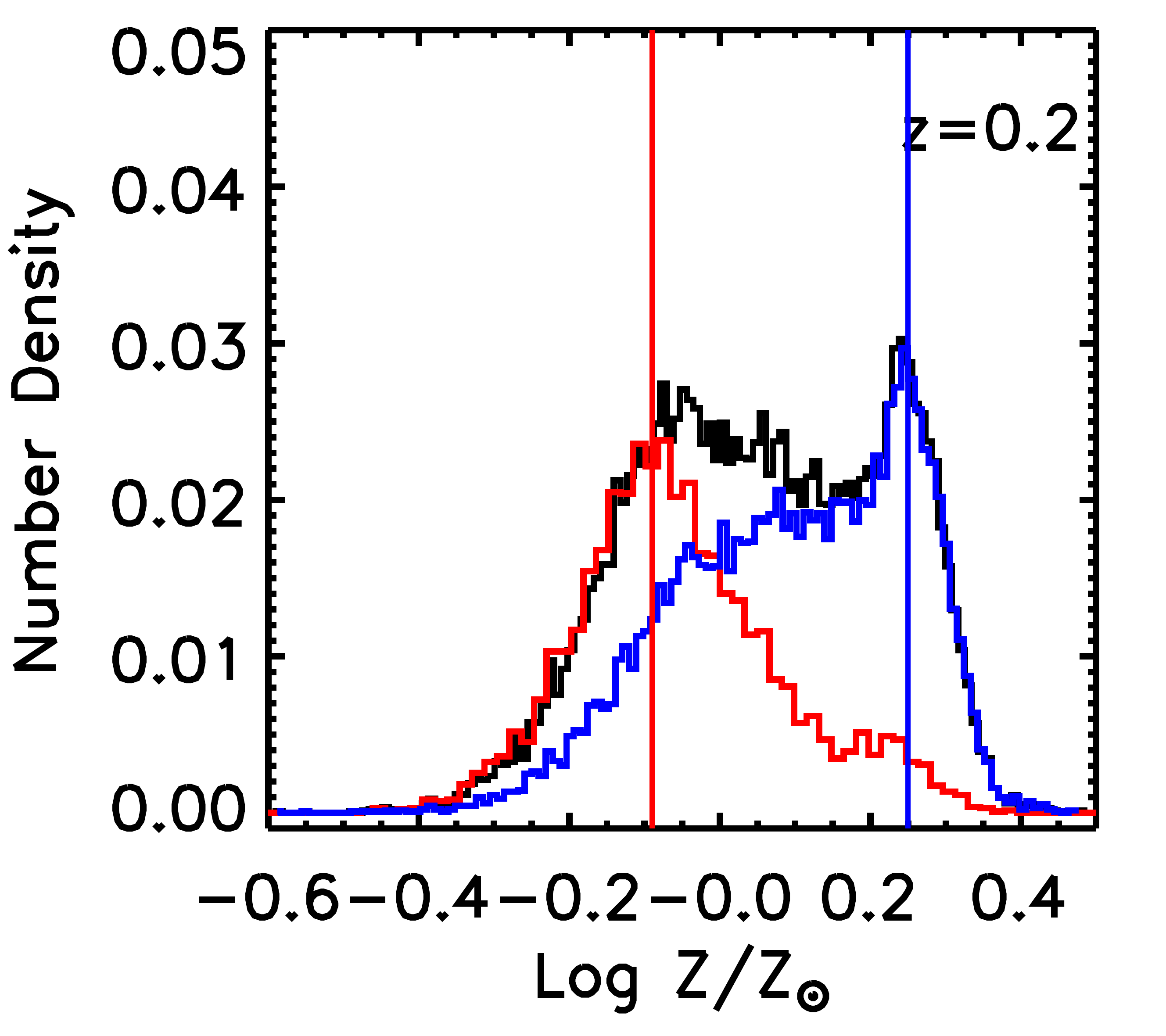}
\includegraphics[width = 0.3\textwidth]{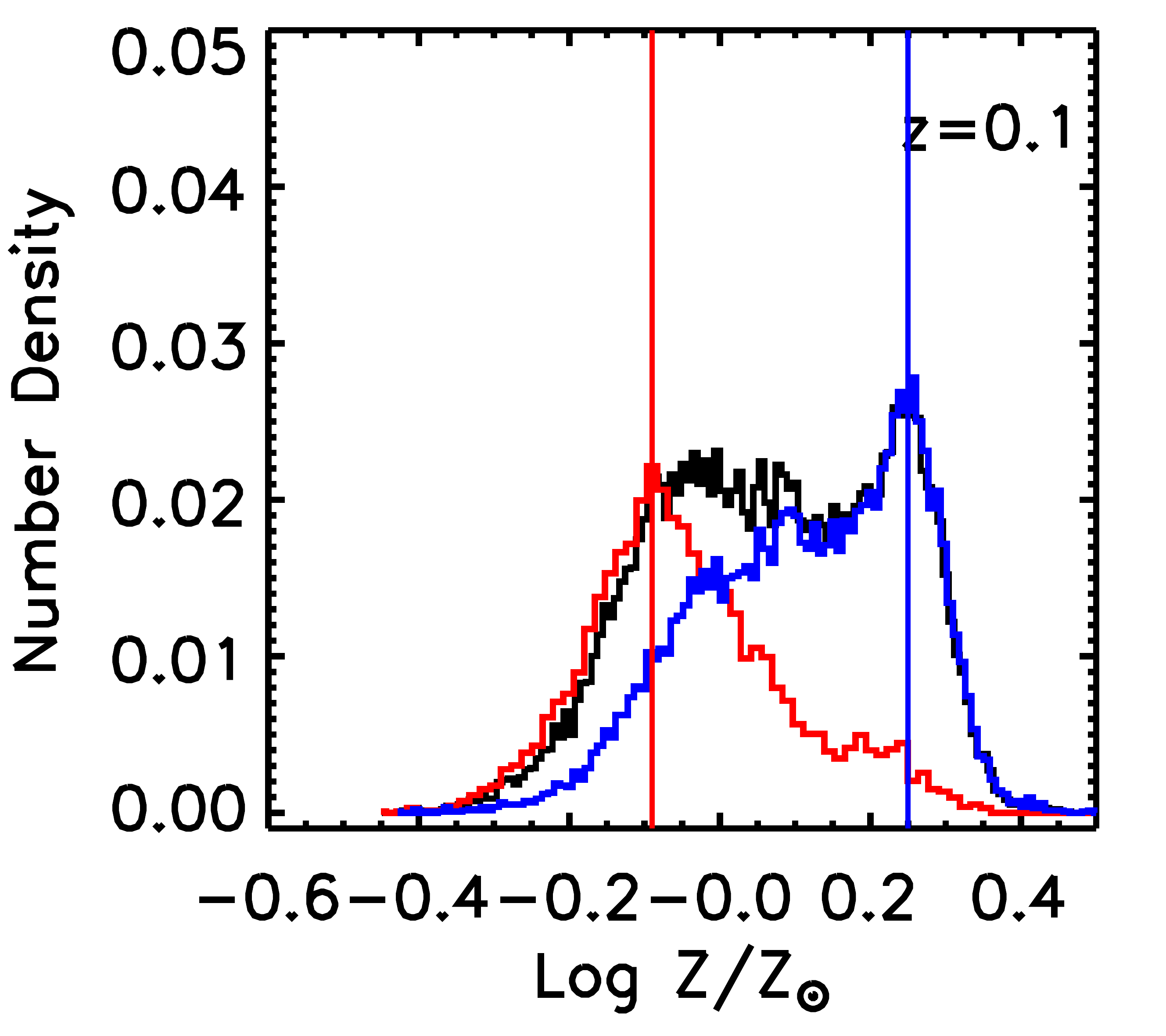}
\caption{
Number density of the stellar metallicities for the full sample (black lines), LSBGs (red lines), HSBGs (blue lines), at redshifts, $z=0.1,\ 0.2,\ 0.3,\ 0.5,\ 1.0,\ 1.5,\ 2.0,\ 3.0,\ 4.0$.
The number density of LSBGs are separated from HSBGs, and the full sample expresses a clear double peaks at redshift $z \le 0.5$.
Similar to Figure ~\ref{Metallicity_distribution}, the bin sizes in each histograms are not same, set as the $autobin$, to obtain a distribution with comparable height.
Note that the number density means galaxy number in a given bin divided by the total number of galaxies.
The vertical lines at each panels represent the peaks of the three kinds of number densities.
The values of these peaks are listed in Table \ref{peak_table}.
We do not plot the number densities for the full sample at low redshifts ($z \le 1$), as they show the double peaks.
}
  \label{Metallicity_distribution_Z}
\end{figure*}
In Figure ~\ref{Metallicity_distribution_Z}, we examine the number densities of stellar metallicities at nine different redshifts, $z=0.1,\ 0.2,\ 0.3,\ 0.5,\ 1.0,\ 1.5,\ 2.0,\ 3.0,\ 4.0$.
Note that the galaxies at $z>0$ are the progenitors of the galaxies at $z=0$ defined through the SUBLINK merger trees \citep{Rodriguez-Gomez2015}.
The black, red, and blue lines in Figure ~\ref{Metallicity_distribution_Z} represent the results of the full sample, LSBGs, and HSBGs, respectively. We list the values of the peaks indicated by the vertical lines with different colours in Table \ref{peak_table}, helping to visualize the peak displacement with redshift.
Note that we apply the $autobin$ for plotting the histograms, which adjusts bin sizes automatically rather than using fixed sizes.
We discuss the results as follows.

At redshifts $z=4.0$ and $3.0$ as shown in the top-left and top-middle panels of Figure ~\ref{Metallicity_distribution_Z}, the number densities for the full sample, LSBGs, and HSBGs exhibit similar peaks around $\log Z/Z_{\odot}\sim -0.80$ and $-0.53$, respectively.
At $z=2.0$ (top-right panel), the number density of HSBGs is higher within the range log $\log Z/Z_{\odot}>0.0$, comparing with that of HSBGs at $z=4$ and $3$. 
Specifically, there is a steep drop at $\log Z/Z_{\odot}\sim0.3$.
At $z=1.5$ (middle-left panel), the number density of HSBGs is almost constant within the range $0.0 < \log Z/Z_{\odot} < 0.4$, but significantly drops at $\log Z/Z_{\odot} > 0.4$.
The peak values for these three samples at $z = 2.0$ and $1.5$ are approximately $\log Z/Z_{\odot}\sim-0.42$ and $-0.32$, respectively.

At redshifts $z\le 1$, the number density of the full sample evidently increases around $\log Z/Z_{\odot}\sim0.25$, emerging a distribution with increasingly prominent double peaks. 
Comparing LSBGs and HSBGs at various redshifts, the peak around $\log Z/Z_{\odot}\sim0.25$ for the full sample primarily arises from the rapid increase in the number of the metal-rich HSBGs.
For LSBGs at redshifts $z=1.0,\ 0.5,\ 0.3,\ 0.2,\ 0.1$, their peak values (indicated by the red vertical lines) are approximately distributed at around $-0.25,\ -0.125, -0.10,\ -0.09,\ -0.09$, whereas the majority of HSBGs are distributed at around $\log Z/Z_{\odot}\sim0.25$ (indicated by the blue vertical lines) at these low redshifts.

The peak value of LSBGs gradually increases from $\log Z/Z_{\odot}\sim-0.80$ to $\sim -0.10$ with decreasing redshifts, indicating an evolutionary trend toward the metal-rich.
Similarly, a large number of HSBGs are metal enriched from $\log Z/Z_{\odot}\sim-0.80$ to $\sim 0.25$.
At higher redshifts ($z > 1.5$), the stellar metal enrichments of both LSBGs and HSBGs appear similar.
However, a distinctive divergence in stellar metallicity evolution becomes evident between LSBGs and HSBGs at lower redshifts ($z < 1.5$).
LSBGs undergo a gradual metal enrichment reaching $\log Z/Z_{\odot}\sim-0.10$, whereas most HSBGs evolve such that they exhibit metallicities higher than solar.
The bifurcated evolution at redshift $z \sim 1.5$ is also found in \cite{Luis2022}, which shows clear bifurcations at $z\sim 0.5-1.5$, $1.5$, and $2.0$, for the evolution of the radius containing half of the total stellar mass ($r_{half}$), the specific angular momentum of the stellar component ($j_{\star}$), and the halo spin parameter ($\lambda$), respectively.
The spin shows the earliest bifurcation in the evolution of the above galaxy properties, as a consequence of being the origin of LSBGs.
The metallicity evolution can be simply associated to the evolution of galaxy size, as the larger galaxies for a given stellar mass (such as LSBGs) are more diffuse, evolve slower, and more metal-poor, whereas the opposite is true for compact galaxies (such as HSBGs).
This supports our result that the increase in the stellar metallicity of LSBGs is slower and more stable than that of HSBGs.
Those findings strongly suggest distinct formation histories and enrichment processes in stellar metallicity between LSBGs and HSBGs.

\section{Conclusions}\label{Conclusions}
In this work, we investigate the stellar metallicity of LSBGs and HSBGs in the IllustrisTNG100-1 simulation.
The LSBG distinction from HSBGs is according to the methodology employed by \cite{Luis2022}.
Our analysis includes the number density and its redshift evolution, radial distribution, and relationships with galaxy properties.
The results we obtained are summarized as the following:
\begin{enumerate}
\item The number distribution of stellar metallicity in galaxies displays a bimodality primarily driven by differing metallicity between LSBGs and HSBGs. 
The metal-poor of LSBGs is a consequence of a radial profile with a sharp decline in stellar metallicity, compared to HSBGs.
\item Initially, the number density of stellar metallicity for galaxies exhibits a single peak during redshifts $z>1.5$. 
As it is evolved, it becomes a bimodal distribution at redshifts $z<1.5$. 
The overall evolutionary of metallicity is consistent with the results in previous works \citep[e.g.,][]{Nelson&Kauffmann2018,Torrey2019}, which suggest that galaxies are more metal-rich at lower redshifts.
\item We find a positive correlation between the stellar metallicity and various properties of galaxies such as half-light radius, stellar mass, and mean age of all the star components.
And, LSBGs are typical more metal-poor, lager, and older than HSBGs.
The metallicity scaling relation is consistent with recent works \citep[e.g.,][]{Torrey2019, Nanni2024, Looser2024}.
\end{enumerate}

The bimodal distribution at lower redshifts is mainly driven by the rapid metal enrichment predominantly seen in HSBGs, dominating higher stellar metallicities. 
Conversely, LSBGs undergo a slower enrichment process, dominating lower stellar metallicities.
This divergence suggests a distinct evolution in the stellar metallicity of LSBGs compared to HSBGs, which gives a new insight on the formation and evolution of LSBGs.


\section*{Acknowledgements}

The author thank the editors and anonymous referee for the useful suggestions and the Illustris and IllustrisTNG projects for providing simulation data.
We acknowledge support from the NSFC grant (Nos. 12003079, 12073089, 12273027) and the National Key Program for Science and Technology Research and Development (No. 2017YFB0203300).
L.T. is supported by Natural Science Foundation of Sichuan Province (No. 2022NSFSC1842), the Fundamental Research Funds of China West Normal University (No. 21E029), and the Sichuan Youth Science and Technology Innovation Research Team (21CXTD0038). 
The analysis carried out in this work is done on the Kunlun HPC facility of the School of Physics and Astronomy, Sun Yat-Sen University.
\section*{Software}
This work is relied on IDL software, including \texttt{IDLAstro} and \texttt{Coyote Graphics Routines}.
The mock observation code may be made available upon reasonable request to the corresponding author.

\section*{ data availability}
IllustrisTNG simulations: \url{https://www.tng-project.org}
The mock galaxy samples or statistical data may be made available upon reasonable request to the corresponding author.

\bibliographystyle{mnras}
\bibliography{main}

\begin{thebibliography}{}
\makeatletter
\relax
\def\mn@urlcharsother{\let\do\@makeother \do\$\do\&\do\#\do\^\do\_\do\%\do\~}
\def\mn@doi{\begingroup\mn@urlcharsother \@ifnextchar [ {\mn@doi@}
  {\mn@doi@[]}}
\def\mn@doi@[#1]#2{\def\@tempa{#1}\ifx\@tempa\@empty \href
  {http://dx.doi.org/#2} {doi:#2}\else \href {http://dx.doi.org/#2} {#1}\fi
  \endgroup}
\def\mn@eprint#1#2{\mn@eprint@#1:#2::\@nil}
\def\mn@eprint@arXiv#1{\href {http://arxiv.org/abs/#1} {{\tt arXiv:#1}}}
\def\mn@eprint@dblp#1{\href {http://dblp.uni-trier.de/rec/bibtex/#1.xml}
  {dblp:#1}}
\def\mn@eprint@#1:#2:#3:#4\@nil{\def\@tempa {#1}\def\@tempb {#2}\def\@tempc
  {#3}\ifx \@tempc \@empty \let \@tempc \@tempb \let \@tempb \@tempa \fi \ifx
  \@tempb \@empty \def\@tempb {arXiv}\fi \@ifundefined
  {mn@eprint@\@tempb}{\@tempb:\@tempc}{\expandafter \expandafter \csname
  mn@eprint@\@tempb\endcsname \expandafter{\@tempc}}}

\bibitem[\protect\citeauthoryear{{Adami}, {Pell{\'o}}, {Ulmer}, {Cuillandre},
  {Durret}, {Mazure}, {Picat}  \& {Scheidegger}}{{Adami}
  et~al.}{2009}]{Adami2009}
{Adami} C.,  {Pell{\'o}} R.,  {Ulmer} M.~P.,  {Cuillandre} J.~C.,  {Durret} F.,
   {Mazure} A.,  {Picat} J.~P.,   {Scheidegger} R.,  2009, \mn@doi [\aap]
  {10.1051/0004-6361:200809858}, \href
  {https://ui.adsabs.harvard.edu/abs/2009A&A...495..407A} {495, 407}

\bibitem[\protect\citeauthoryear{{Allen} \& {Shu}}{{Allen} \&
  {Shu}}{1979}]{Allen1979}
{Allen} R.~J.,  {Shu} F.~H.,  1979, \mn@doi [\apj] {10.1086/156705}, \href
  {https://ui.adsabs.harvard.edu/abs/1979ApJ...227...67A} {227, 67}

\bibitem[\protect\citeauthoryear{{Auld}, {de Blok}, {Bell}  \& {Davies}}{{Auld}
  et~al.}{2006}]{Auld2006}
{Auld} R.,  {de Blok} W.~J.~G.,  {Bell} E.,   {Davies} J.~I.,  2006, \mn@doi
  [\mnras] {10.1111/j.1365-2966.2005.09924.x}, \href
  {https://ui.adsabs.harvard.edu/abs/2006MNRAS.366.1475A} {366, 1475}

\bibitem[\protect\citeauthoryear{{Baldry}, {Glazebrook}, {Brinkmann},
  {Ivezi{\'c}}, {Lupton}, {Nichol}  \& {Szalay}}{{Baldry}
  et~al.}{2004}]{Baldry2004}
{Baldry} I.~K.,  {Glazebrook} K.,  {Brinkmann} J.,  {Ivezi{\'c}} {\v{Z}}.,
  {Lupton} R.~H.,  {Nichol} R.~C.,   {Szalay} A.~S.,  2004, \mn@doi [\apj]
  {10.1086/380092}, \href
  {https://ui.adsabs.harvard.edu/abs/2004ApJ...600..681B} {600, 681}

\bibitem[\protect\citeauthoryear{{Behroozi}, {Wechsler}  \&
  {Conroy}}{{Behroozi} et~al.}{2013}]{Behroozi2013}
{Behroozi} P.~S.,  {Wechsler} R.~H.,   {Conroy} C.,  2013, \mn@doi [\apj]
  {10.1088/0004-637X/770/1/57}, \href
  {https://ui.adsabs.harvard.edu/abs/2013ApJ...770...57B} {770, 57}

\bibitem[\protect\citeauthoryear{{Benavides} et~al.,}{{Benavides}
  et~al.}{2021}]{Benavides2021}
{Benavides} J.~A.,  et~al., 2021, \mn@doi [Nature Astronomy]
  {10.1038/s41550-021-01458-1}, \href
  {https://ui.adsabs.harvard.edu/abs/2021NatAs...5.1255B} {5, 1255}

\bibitem[\protect\citeauthoryear{{Bothun}, {Impey}  \& {McGaugh}}{{Bothun}
  et~al.}{1997}]{Bothun1997}
{Bothun} G.,  {Impey} C.,   {McGaugh} S.,  1997, \mn@doi [\pasp]
  {10.1086/133941}, \href
  {https://ui.adsabs.harvard.edu/abs/1997PASP..109..745B} {109, 745}

\bibitem[\protect\citeauthoryear{{Cao}, {Wu}, {Galaz}, {Kalari}, {Cheng}, {Li}
  \& {Wang}}{{Cao} et~al.}{2023}]{Cao2023}
{Cao} T.-w.,  {Wu} H.,  {Galaz} G.,  {Kalari} V.~M.,  {Cheng} C.,  {Li} Z.-J.,
   {Wang} J.-f.,  2023, \mn@doi [\apj] {10.3847/1538-4357/acc864}, \href
  {https://ui.adsabs.harvard.edu/abs/2023ApJ...948...96C} {948, 96}

\bibitem[\protect\citeauthoryear{{Dalcanton}, {Spergel}  \&
  {Summers}}{{Dalcanton} et~al.}{1997}]{Dalcanton1997}
{Dalcanton} J.~J.,  {Spergel} D.~N.,   {Summers} F.~J.,  1997, \mn@doi [\apj]
  {10.1086/304182}, \href
  {https://ui.adsabs.harvard.edu/abs/1997ApJ...482..659D} {482, 659}

\bibitem[\protect\citeauthoryear{{Davis}, {Efstathiou}, {Frenk}  \&
  {White}}{{Davis} et~al.}{1985}]{Davis1985}
{Davis} M.,  {Efstathiou} G.,  {Frenk} C.~S.,   {White} S.~D.~M.,  1985,
  \mn@doi [\apj] {10.1086/163168}, \href
  {https://ui.adsabs.harvard.edu/abs/1985ApJ...292..371D} {292, 371}

\bibitem[\protect\citeauthoryear{{Dekel} \& {Birnboim}}{{Dekel} \&
  {Birnboim}}{2006}]{Dekel2006}
{Dekel} A.,  {Birnboim} Y.,  2006, \mn@doi [\mnras]
  {10.1111/j.1365-2966.2006.10145.x}, \href
  {https://ui.adsabs.harvard.edu/abs/2006MNRAS.368....2D} {368, 2}

\bibitem[\protect\citeauthoryear{{Di Cintio}, {Brook}, {Dutton}, {Macci{\`o}},
  {Obreja}  \& {Dekel}}{{Di Cintio} et~al.}{2017}]{DiCintio2017}
{Di Cintio} A.,  {Brook} C.~B.,  {Dutton} A.~A.,  {Macci{\`o}} A.~V.,  {Obreja}
  A.,   {Dekel} A.,  2017, \mn@doi [\mnras] {10.1093/mnrasl/slw210}, \href
  {https://ui.adsabs.harvard.edu/abs/2017MNRAS.466L...1D} {466, L1}

\bibitem[\protect\citeauthoryear{{Di Cintio}, {Brook}, {Macci{\`o}}, {Dutton}
  \& {Cardona-Barrero}}{{Di Cintio} et~al.}{2019}]{DiCintio2019}
{Di Cintio} A.,  {Brook} C.~B.,  {Macci{\`o}} A.~V.,  {Dutton} A.~A.,
  {Cardona-Barrero} S.,  2019, \mn@doi [\mnras] {10.1093/mnras/stz985}, \href
  {https://ui.adsabs.harvard.edu/abs/2019MNRAS.486.2535D} {486, 2535}

\bibitem[\protect\citeauthoryear{{Disney}}{{Disney}}{1976}]{Disney1976}
{Disney} M.~J.,  1976, \mn@doi [\nat] {10.1038/263573a0}, \href
  {https://ui.adsabs.harvard.edu/abs/1976Natur.263..573D} {263, 573}

\bibitem[\protect\citeauthoryear{{Dolag}, {Borgani}, {Murante}  \&
  {Springel}}{{Dolag} et~al.}{2009}]{Dolag2009}
{Dolag} K.,  {Borgani} S.,  {Murante} G.,   {Springel} V.,  2009, \mn@doi
  [\mnras] {10.1111/j.1365-2966.2009.15034.x}, \href
  {https://ui.adsabs.harvard.edu/abs/2009MNRAS.399..497D} {399, 497}

\bibitem[\protect\citeauthoryear{{Du}, {Wu}, {Lam}, {Zhu}, {Lei}  \&
  {Zhou}}{{Du} et~al.}{2015}]{Du2015}
{Du} W.,  {Wu} H.,  {Lam} M.~I.,  {Zhu} Y.,  {Lei} F.,   {Zhou} Z.,  2015,
  \mn@doi [\aj] {10.1088/0004-6256/149/6/199}, \href
  {https://ui.adsabs.harvard.edu/abs/2015AJ....149..199D} {149, 199}

\bibitem[\protect\citeauthoryear{{Dubois} et~al.,}{{Dubois}
  et~al.}{2014}]{Dubois2014}
{Dubois} Y.,  et~al., 2014, \mn@doi [\mnras] {10.1093/mnras/stu1227}, \href
  {https://ui.adsabs.harvard.edu/abs/2014MNRAS.444.1453D} {444, 1453}

\bibitem[\protect\citeauthoryear{{Freeman}}{{Freeman}}{1970}]{Freeman1970}
{Freeman} K.~C.,  1970, \mn@doi [\apj] {10.1086/150474}, \href
  {https://ui.adsabs.harvard.edu/abs/1970ApJ...160..811F} {160, 811}

\bibitem[\protect\citeauthoryear{{Fu} et~al.,}{{Fu} et~al.}{2023}]{Fu2023}
{Fu} J.,  et~al., 2023, \mn@doi [arXiv e-prints] {10.48550/arXiv.2311.05621},
  \href {https://ui.adsabs.harvard.edu/abs/2023arXiv231105621F} {p.
  arXiv:2311.05621}

\bibitem[\protect\citeauthoryear{{Galaz}, {Villalobos}, {Infante}  \&
  {Donzelli}}{{Galaz} et~al.}{2006}]{Galaz2006}
{Galaz} G.,  {Villalobos} A.,  {Infante} L.,   {Donzelli} C.,  2006, \mn@doi
  [\aj] {10.1086/500931}, \href
  {https://ui.adsabs.harvard.edu/abs/2006AJ....131.2035G} {131, 2035}

\bibitem[\protect\citeauthoryear{{Gonz{\'a}lez Delgado} et~al.,}{{Gonz{\'a}lez
  Delgado} et~al.}{2014}]{Delgado2014}
{Gonz{\'a}lez Delgado} R.~M.,  et~al., 2014, \mn@doi [\apjl]
  {10.1088/2041-8205/791/1/L16}, \href
  {https://ui.adsabs.harvard.edu/abs/2014ApJ...791L..16G} {791, L16}

\bibitem[\protect\citeauthoryear{{Haberzettl}, {Bomans}  \&
  {Dettmar}}{{Haberzettl} et~al.}{2007}]{Haberzettl2007}
{Haberzettl} L.,  {Bomans} D.~J.,   {Dettmar} R.~J.,  2007, \mn@doi [\aap]
  {10.1051/0004-6361:20066918}, \href
  {https://ui.adsabs.harvard.edu/abs/2007A&A...471..787H} {471, 787}

\bibitem[\protect\citeauthoryear{{Hayward}, {Irwin}  \& {Bregman}}{{Hayward}
  et~al.}{2005}]{Hayward2005}
{Hayward} C.~C.,  {Irwin} J.~A.,   {Bregman} J.~N.,  2005, \mn@doi [\apj]
  {10.1086/497565}, \href
  {https://ui.adsabs.harvard.edu/abs/2005ApJ...635..827H} {635, 827}

\bibitem[\protect\citeauthoryear{{Heller} \& {Brosch}}{{Heller} \&
  {Brosch}}{2001}]{Heller2001}
{Heller} A.~B.,  {Brosch} N.,  2001, \mn@doi [\mnras]
  {10.1046/j.1365-8711.2001.04636.x}, \href
  {https://ui.adsabs.harvard.edu/abs/2001MNRAS.327...80H} {327, 80}

\bibitem[\protect\citeauthoryear{{Hemler} et~al.,}{{Hemler}
  et~al.}{2021}]{Hemler2021}
{Hemler} Z.~S.,  et~al., 2021, \mn@doi [\mnras] {10.1093/mnras/stab1803}, \href
  {https://ui.adsabs.harvard.edu/abs/2021MNRAS.506.3024H} {506, 3024}

\bibitem[\protect\citeauthoryear{{Huang} et~al.,}{{Huang}
  et~al.}{2014}]{Huang2014}
{Huang} S.,  et~al., 2014, \mn@doi [\apj] {10.1088/0004-637X/793/1/40}, \href
  {https://ui.adsabs.harvard.edu/abs/2014ApJ...793...40H} {793, 40}

\bibitem[\protect\citeauthoryear{{Impey} \& {Bothun}}{{Impey} \&
  {Bothun}}{1997}]{Impey1997}
{Impey} C.,  {Bothun} G.,  1997, \mn@doi [\araa]
  {10.1146/annurev.astro.35.1.267}, \href
  {https://ui.adsabs.harvard.edu/abs/1997ARA&A..35..267I} {35, 267}

\bibitem[\protect\citeauthoryear{{Jimenez}, {Padoan}, {Matteucci}  \&
  {Heavens}}{{Jimenez} et~al.}{1998}]{Jimenez1998}
{Jimenez} R.,  {Padoan} P.,  {Matteucci} F.,   {Heavens} A.~F.,  1998, \mn@doi
  [\mnras] {10.1046/j.1365-8711.1998.01731.x}, \href
  {https://ui.adsabs.harvard.edu/abs/1998MNRAS.299..123J} {299, 123}

\bibitem[\protect\citeauthoryear{{Junais} et~al.,}{{Junais}
  et~al.}{2023}]{Junais2023}
{Junais} et~al., 2023, \mn@doi [arXiv e-prints] {10.48550/arXiv.2310.11872},
  \href {https://ui.adsabs.harvard.edu/abs/2023arXiv231011872J} {p.
  arXiv:2310.11872}

\bibitem[\protect\citeauthoryear{{Kim} \& {Lee}}{{Kim} \&
  {Lee}}{2013}]{Kim&Lee2013}
{Kim} J.-h.,  {Lee} J.,  2013, \mn@doi [\mnras] {10.1093/mnras/stt632}, \href
  {https://ui.adsabs.harvard.edu/abs/2013MNRAS.432.1701K} {432, 1701}

\bibitem[\protect\citeauthoryear{{Kim} \& {McGaugh}}{{Kim} \&
  {McGaugh}}{2012}]{Kim2012}
{Kim} J.~H.,  {McGaugh} S.,  2012, in {Aoki} W.,  {Ishigaki} M.,  {Suda} T.,
  {Tsujimoto} T.,   {Arimoto} N.,  eds,  Astronomical Society of the Pacific
  Conference Series Vol. 458, Galactic Archaeology: Near-Field Cosmology and
  the Formation of the Milky Way. p.~327

\bibitem[\protect\citeauthoryear{{Koda}, {Yagi}, {Yamanoi}  \&
  {Komiyama}}{{Koda} et~al.}{2015}]{Koda2015}
{Koda} J.,  {Yagi} M.,  {Yamanoi} H.,   {Komiyama} Y.,  2015, \mn@doi [\apjl]
  {10.1088/2041-8205/807/1/L2}, \href
  {https://ui.adsabs.harvard.edu/abs/2015ApJ...807L...2K} {807, L2}

\bibitem[\protect\citeauthoryear{{Kulier}, {Galaz}, {Padilla}  \&
  {Trayford}}{{Kulier} et~al.}{2020}]{Kulier2020}
{Kulier} A.,  {Galaz} G.,  {Padilla} N.~D.,   {Trayford} J.~W.,  2020, \mn@doi
  [\mnras] {10.1093/mnras/staa1798}, \href
  {https://ui.adsabs.harvard.edu/abs/2020MNRAS.496.3996K} {496, 3996}

\bibitem[\protect\citeauthoryear{{Kuzio de Naray}, {McGaugh}  \& {de
  Blok}}{{Kuzio de Naray} et~al.}{2008}]{Kuzio2008}
{Kuzio de Naray} R.,  {McGaugh} S.~S.,   {de Blok} W.~J.~G.,  2008, \mn@doi
  [\apj] {10.1086/527543}, \href
  {https://ui.adsabs.harvard.edu/abs/2008ApJ...676..920K} {676, 920}

\bibitem[\protect\citeauthoryear{{Liang} et~al.,}{{Liang}
  et~al.}{2010}]{Liang2010}
{Liang} Y.~C.,  et~al., 2010, \mn@doi [\mnras]
  {10.1111/j.1365-2966.2010.16891.x}, \href
  {https://ui.adsabs.harvard.edu/abs/2010MNRAS.409..213L} {409, 213}

\bibitem[\protect\citeauthoryear{{Looser}, {D'Eugenio}, {Piotrowska},
  {Belfiore}, {Maiolino}, {Cappellari}, {Baker}  \& {Tacchella}}{{Looser}
  et~al.}{2024}]{Looser2024}
{Looser} T.~J.,  {D'Eugenio} F.,  {Piotrowska} J.~M.,  {Belfiore} F.,
  {Maiolino} R.,  {Cappellari} M.,  {Baker} W.~M.,   {Tacchella} S.,  2024,
  \mn@doi [arXiv e-prints] {10.48550/arXiv.2401.08769}, \href
  {https://ui.adsabs.harvard.edu/abs/2024arXiv240108769L} {p. arXiv:2401.08769}

\bibitem[\protect\citeauthoryear{{Marinacci} et~al.,}{{Marinacci}
  et~al.}{2018}]{Marinacci2018}
{Marinacci} F.,  et~al., 2018, \mn@doi [\mnras] {10.1093/mnras/sty2206}, \href
  {https://ui.adsabs.harvard.edu/abs/2018MNRAS.480.5113M} {480, 5113}

\bibitem[\protect\citeauthoryear{{Martin} et~al.,}{{Martin}
  et~al.}{2019}]{Martin2019}
{Martin} G.,  et~al., 2019, \mn@doi [\mnras] {10.1093/mnras/stz356}, \href
  {https://ui.adsabs.harvard.edu/abs/2019MNRAS.485..796M} {485, 796}

\bibitem[\protect\citeauthoryear{{McGaugh}}{{McGaugh}}{1994}]{McGaugh1994}
{McGaugh} S.~S.,  1994, \mn@doi [\apj] {10.1086/174049}, \href
  {https://ui.adsabs.harvard.edu/abs/1994ApJ...426..135M} {426, 135}

\bibitem[\protect\citeauthoryear{{McGaugh}}{{McGaugh}}{2021}]{McGaugh2021}
{McGaugh} S.~S.,  2021, \mn@doi [Studies in History and Philosophy of Science]
  {10.1016/j.shpsa.2021.05.008}, \href
  {https://ui.adsabs.harvard.edu/abs/2021SHPSA..88..220M} {88, 220}

\bibitem[\protect\citeauthoryear{{McGaugh}, {Schombert}  \& {Bothun}}{{McGaugh}
  et~al.}{1995}]{McGaugh1995b}
{McGaugh} S.~S.,  {Schombert} J.~M.,   {Bothun} G.~D.,  1995, \mn@doi [\aj]
  {10.1086/117427}, \href
  {https://ui.adsabs.harvard.edu/abs/1995AJ....109.2019M} {109, 2019}

\bibitem[\protect\citeauthoryear{{Merritt}, {van Dokkum}, {Danieli}, {Abraham},
  {Zhang}, {Karachentsev}  \& {Makarova}}{{Merritt} et~al.}{2016}]{Merritt2016}
{Merritt} A.,  {van Dokkum} P.,  {Danieli} S.,  {Abraham} R.,  {Zhang} J.,
  {Karachentsev} I.~D.,   {Makarova} L.~N.,  2016, \mn@doi [\apj]
  {10.3847/1538-4357/833/2/168}, \href
  {https://ui.adsabs.harvard.edu/abs/2016ApJ...833..168M} {833, 168}

\bibitem[\protect\citeauthoryear{{Minchin} et~al.,}{{Minchin}
  et~al.}{2004}]{Minchin2004}
{Minchin} R.~F.,  et~al., 2004, \mn@doi [\mnras]
  {10.1111/j.1365-2966.2004.08409.x}, \href
  {https://ui.adsabs.harvard.edu/abs/2004MNRAS.355.1303M} {355, 1303}

\bibitem[\protect\citeauthoryear{{Naiman} et~al.,}{{Naiman}
  et~al.}{2018}]{Naiman2018}
{Naiman} J.~P.,  et~al., 2018, \mn@doi [\mnras] {10.1093/mnras/sty618}, \href
  {https://ui.adsabs.harvard.edu/abs/2018MNRAS.477.1206N} {477, 1206}

\bibitem[\protect\citeauthoryear{{Nanni} et~al.,}{{Nanni}
  et~al.}{2024}]{Nanni2024}
{Nanni} L.,  et~al., 2024, \mn@doi [\mnras] {10.1093/mnras/stad3599}, \href
  {https://ui.adsabs.harvard.edu/abs/2024MNRAS.527.6419N} {527, 6419}

\bibitem[\protect\citeauthoryear{{Nelson} et~al.,}{{Nelson}
  et~al.}{2018a}]{Nelson2018}
{Nelson} D.,  et~al., 2018a, \mn@doi [\mnras] {10.1093/mnras/stx3040}, \href
  {https://ui.adsabs.harvard.edu/abs/2018MNRAS.475..624N} {475, 624}

\bibitem[\protect\citeauthoryear{{Nelson} et~al.,}{{Nelson}
  et~al.}{2018b}]{Nelson&Kauffmann2018}
{Nelson} D.,  et~al., 2018b, \mn@doi [\mnras] {10.1093/mnras/sty656}, \href
  {https://ui.adsabs.harvard.edu/abs/2018MNRAS.477..450N} {477, 450}

\bibitem[\protect\citeauthoryear{{Nelson} et~al.,}{{Nelson}
  et~al.}{2019}]{Nelson2019}
{Nelson} D.,  et~al., 2019, \mn@doi [Computational Astrophysics and Cosmology]
  {10.1186/s40668-019-0028-x}, \href
  {https://ui.adsabs.harvard.edu/abs/2019ComAC...6....2N} {6, 2}

\bibitem[\protect\citeauthoryear{{O'Neil}, {Bothun}, {Schombert}, {Cornell}  \&
  {Impey}}{{O'Neil} et~al.}{1997}]{O'Neil1997}
{O'Neil} K.,  {Bothun} G.~D.,  {Schombert} J.,  {Cornell} M.~E.,   {Impey}
  C.~D.,  1997, \mn@doi [\aj] {10.1086/118659}, \href
  {https://ui.adsabs.harvard.edu/abs/1997AJ....114.2448O} {114, 2448}

\bibitem[\protect\citeauthoryear{{P{\'e}rez-Monta{\~n}o} \& {Cervantes
  Sodi}}{{P{\'e}rez-Monta{\~n}o} \& {Cervantes
  Sodi}}{2019}]{Enrique&Bernardo2019}
{P{\'e}rez-Monta{\~n}o} L.~E.,  {Cervantes Sodi} B.,  2019, \mn@doi [\mnras]
  {10.1093/mnras/stz2847}, \href
  {https://ui.adsabs.harvard.edu/abs/2019MNRAS.490.3772P} {490, 3772}

\bibitem[\protect\citeauthoryear{{P{\'e}rez-Monta{\~n}o}, {Rodriguez-Gomez},
  {Cervantes Sodi}, {Zhu}, {Pillepich}, {Vogelsberger}  \&
  {Hernquist}}{{P{\'e}rez-Monta{\~n}o} et~al.}{2022}]{Luis2022}
{P{\'e}rez-Monta{\~n}o} L.~E.,  {Rodriguez-Gomez} V.,  {Cervantes Sodi} B.,
  {Zhu} Q.,  {Pillepich} A.,  {Vogelsberger} M.,   {Hernquist} L.,  2022,
  \mn@doi [\mnras] {10.1093/mnras/stac1716}, \href
  {https://ui.adsabs.harvard.edu/abs/2022MNRAS.514.5840P} {514, 5840}

\bibitem[\protect\citeauthoryear{{Pillepich} et~al.,}{{Pillepich}
  et~al.}{2018}]{Pillepich2018}
{Pillepich} A.,  et~al., 2018, \mn@doi [\mnras] {10.1093/mnras/stx2656}, \href
  {https://ui.adsabs.harvard.edu/abs/2018MNRAS.473.4077P} {473, 4077}

\bibitem[\protect\citeauthoryear{{Rodriguez-Gomez} et~al.,}{{Rodriguez-Gomez}
  et~al.}{2015}]{Rodriguez-Gomez2015}
{Rodriguez-Gomez} V.,  et~al., 2015, \mn@doi [\mnras] {10.1093/mnras/stv264},
  \href {https://ui.adsabs.harvard.edu/abs/2015MNRAS.449...49R} {449, 49}

\bibitem[\protect\citeauthoryear{{Rom{\'a}n} \& {Trujillo}}{{Rom{\'a}n} \&
  {Trujillo}}{2017}]{Roman2017}
{Rom{\'a}n} J.,  {Trujillo} I.,  2017, \mn@doi [\mnras] {10.1093/mnras/stx694},
  \href {https://ui.adsabs.harvard.edu/abs/2017MNRAS.468.4039R} {468, 4039}

\bibitem[\protect\citeauthoryear{{Ruiz-Lara} et~al.,}{{Ruiz-Lara}
  et~al.}{2018}]{Ruiz-Lara2018}
{Ruiz-Lara} T.,  et~al., 2018, \mn@doi [\mnras] {10.1093/mnras/sty1112}, \href
  {https://ui.adsabs.harvard.edu/abs/2018MNRAS.478.2034R} {478, 2034}

\bibitem[\protect\citeauthoryear{{Sales}, {Navarro}, {Pe{\~n}afiel}, {Peng},
  {Lim}  \& {Hernquist}}{{Sales} et~al.}{2020}]{Sales2020}
{Sales} L.~V.,  {Navarro} J.~F.,  {Pe{\~n}afiel} L.,  {Peng} E.~W.,  {Lim} S.,
   {Hernquist} L.,  2020, \mn@doi [\mnras] {10.1093/mnras/staa854}, \href
  {https://ui.adsabs.harvard.edu/abs/2020MNRAS.494.1848S} {494, 1848}

\bibitem[\protect\citeauthoryear{{Salinas} \& {Galaz}}{{Salinas} \&
  {Galaz}}{2021}]{Salinas&Galaz2021}
{Salinas} V.~H.,  {Galaz} G.,  2021, \mn@doi [\apj] {10.3847/1538-4357/ac043d},
  \href {https://ui.adsabs.harvard.edu/abs/2021ApJ...915..125S} {915, 125}

\bibitem[\protect\citeauthoryear{{Schaye} et~al.,}{{Schaye}
  et~al.}{2015}]{Schaye2015}
{Schaye} J.,  et~al., 2015, \mn@doi [\mnras] {10.1093/mnras/stu2058}, \href
  {https://ui.adsabs.harvard.edu/abs/2015MNRAS.446..521S} {446, 521}

\bibitem[\protect\citeauthoryear{{Schombert} \& {McGaugh}}{{Schombert} \&
  {McGaugh}}{2021}]{Schombert2021}
{Schombert} J.,  {McGaugh} S.,  2021, \mn@doi [\aj] {10.3847/1538-3881/abd54d},
  \href {https://ui.adsabs.harvard.edu/abs/2021AJ....161...91S} {161, 91}

\bibitem[\protect\citeauthoryear{{Shao}, {Disseau}, {Yang}, {Hammer}, {Puech},
  {Rodrigues}, {Liang}  \& {Deng}}{{Shao} et~al.}{2015}]{Shao2015}
{Shao} X.,  {Disseau} K.,  {Yang} Y.~B.,  {Hammer} F.,  {Puech} M.,
  {Rodrigues} M.,  {Liang} Y.~C.,   {Deng} L.~C.,  2015, \mn@doi [\aap]
  {10.1051/0004-6361/201525796}, \href
  {https://ui.adsabs.harvard.edu/abs/2015A&A...579A..57S} {579, A57}

\bibitem[\protect\citeauthoryear{{Shen}, {Mo}, {White}, {Blanton}, {Kauffmann},
  {Voges}, {Brinkmann}  \& {Csabai}}{{Shen} et~al.}{2003}]{Shen2003}
{Shen} S.,  {Mo} H.~J.,  {White} S. D.~M.,  {Blanton} M.~R.,  {Kauffmann} G.,
  {Voges} W.,  {Brinkmann} J.,   {Csabai} I.,  2003, \mn@doi [\mnras]
  {10.1046/j.1365-8711.2003.06740.x}, \href
  {https://ui.adsabs.harvard.edu/abs/2003MNRAS.343..978S} {343, 978}

\bibitem[\protect\citeauthoryear{{Springel}, {White}, {Tormen}  \&
  {Kauffmann}}{{Springel} et~al.}{2001}]{Springel2001}
{Springel} V.,  {White} S. D.~M.,  {Tormen} G.,   {Kauffmann} G.,  2001,
  \mn@doi [\mnras] {10.1046/j.1365-8711.2001.04912.x}, \href
  {https://ui.adsabs.harvard.edu/abs/2001MNRAS.328..726S} {328, 726}

\bibitem[\protect\citeauthoryear{{Springel} et~al.,}{{Springel}
  et~al.}{2018}]{Springel2018}
{Springel} V.,  et~al., 2018, \mn@doi [\mnras] {10.1093/mnras/stx3304}, \href
  {https://ui.adsabs.harvard.edu/abs/2018MNRAS.475..676S} {475, 676}

\bibitem[\protect\citeauthoryear{{Swaters}, {Madore}, {van den Bosch}  \&
  {Balcells}}{{Swaters} et~al.}{2003}]{Swaters2003}
{Swaters} R.~A.,  {Madore} B.~F.,  {van den Bosch} F.~C.,   {Balcells} M.,
  2003, \mn@doi [\apj] {10.1086/345426}, \href
  {https://ui.adsabs.harvard.edu/abs/2003ApJ...583..732S} {583, 732}

\bibitem[\protect\citeauthoryear{{Torrey} et~al.,}{{Torrey}
  et~al.}{2019}]{Torrey2019}
{Torrey} P.,  et~al., 2019, \mn@doi [\mnras] {10.1093/mnras/stz243}, \href
  {https://ui.adsabs.harvard.edu/abs/2019MNRAS.484.5587T} {484, 5587}

\bibitem[\protect\citeauthoryear{{Vogelsberger} et~al.,}{{Vogelsberger}
  et~al.}{2014}]{Vogelsberger2014}
{Vogelsberger} M.,  et~al., 2014, \mn@doi [\mnras] {10.1093/mnras/stu1536},
  \href {https://ui.adsabs.harvard.edu/abs/2014MNRAS.444.1518V} {444, 1518}

\bibitem[\protect\citeauthoryear{{Wang}, {Dutton}, {Stinson}, {Macci{\`o}},
  {Penzo}, {Kang}, {Keller}  \& {Wadsley}}{{Wang} et~al.}{2015}]{Wang2015}
{Wang} L.,  {Dutton} A.~A.,  {Stinson} G.~S.,  {Macci{\`o}} A.~V.,  {Penzo} C.,
   {Kang} X.,  {Keller} B.~W.,   {Wadsley} J.,  2015, \mn@doi [\mnras]
  {10.1093/mnras/stv1937}, \href
  {https://ui.adsabs.harvard.edu/abs/2015MNRAS.454...83W} {454, 83}

\bibitem[\protect\citeauthoryear{{Weinberger} et~al.,}{{Weinberger}
  et~al.}{2017}]{Weinberger2017}
{Weinberger} R.,  et~al., 2017, \mn@doi [\mnras] {10.1093/mnras/stw2944}, \href
  {https://ui.adsabs.harvard.edu/abs/2017MNRAS.465.3291W} {465, 3291}

\bibitem[\protect\citeauthoryear{{Wright}, {Tremmel}, {Brooks}, {Munshi},
  {Nagai}, {Sharma}  \& {Quinn}}{{Wright} et~al.}{2021}]{Wright2021}
{Wright} A.~C.,  {Tremmel} M.,  {Brooks} A.~M.,  {Munshi} F.,  {Nagai} D.,
  {Sharma} R.~S.,   {Quinn} T.~R.,  2021, \mn@doi [\mnras]
  {10.1093/mnras/stab081}, \href
  {https://ui.adsabs.harvard.edu/abs/2021MNRAS.502.5370W} {502, 5370}

\bibitem[\protect\citeauthoryear{{Young}, {Kuzio de Naray}  \& {Wang}}{{Young}
  et~al.}{2015}]{Young2015}
{Young} J.~E.,  {Kuzio de Naray} R.,   {Wang} S.~X.,  2015, \mn@doi [\mnras]
  {10.1093/mnras/stv1492}, \href
  {https://ui.adsabs.harvard.edu/abs/2015MNRAS.452.2973Y} {452, 2973}

\bibitem[\protect\citeauthoryear{{Zhong}, {Liang}, {Liu}, {Hammer}, {Hu},
  {Chen}, {Deng}  \& {Zhang}}{{Zhong} et~al.}{2008}]{Zhong2008}
{Zhong} G.~H.,  {Liang} Y.~C.,  {Liu} F.~S.,  {Hammer} F.,  {Hu} J.~Y.,  {Chen}
  X.~Y.,  {Deng} L.~C.,   {Zhang} B.,  2008, \mn@doi [\mnras]
  {10.1111/j.1365-2966.2008.13972.x}, \href
  {https://ui.adsabs.harvard.edu/abs/2008MNRAS.391..986Z} {391, 986}

\bibitem[\protect\citeauthoryear{{Zhu} et~al.,}{{Zhu} et~al.}{2018}]{Zhu2018}
{Zhu} Q.,  et~al., 2018, \mn@doi [\mnras] {10.1093/mnrasl/sly111}, \href
  {https://ui.adsabs.harvard.edu/abs/2018MNRAS.480L..18Z} {480, L18}

\bibitem[\protect\citeauthoryear{{Zhu}, {P{\'e}rez-Monta{\~n}o},
  {Rodriguez-Gomez}, {Cervantes Sodi}, {Zjupa}, {Marinacci}, {Vogelsberger}  \&
  {Hernquist}}{{Zhu} et~al.}{2023}]{Zhu2023}
{Zhu} Q.,  {P{\'e}rez-Monta{\~n}o} L.~E.,  {Rodriguez-Gomez} V.,  {Cervantes
  Sodi} B.,  {Zjupa} J.,  {Marinacci} F.,  {Vogelsberger} M.,   {Hernquist} L.,
   2023, \mn@doi [\mnras] {10.1093/mnras/stad1655}, \href
  {https://ui.adsabs.harvard.edu/abs/2023MNRAS.523.3991Z} {523, 3991}

\bibitem[\protect\citeauthoryear{{de Blok}, {van der Hulst}  \& {Bothun}}{{de
  Blok} et~al.}{1995}]{deBlok1995}
{de Blok} W.~J.~G.,  {van der Hulst} J.~M.,   {Bothun} G.~D.,  1995, \mn@doi
  [\mnras] {10.1093/mnras/274.1.235}, \href
  {https://ui.adsabs.harvard.edu/abs/1995MNRAS.274..235D} {274, 235}

\bibitem[\protect\citeauthoryear{{de Blok}, {Walter}  \& {Bell}}{{de Blok}
  et~al.}{1999}]{deBlok1999}
{de Blok} E.,  {Walter} F.,   {Bell} E.,  1999, \mn@doi [\apss]
  {10.1023/A:1017063705721}, \href
  {https://ui.adsabs.harvard.edu/abs/1999Ap&SS.269..101D} {269, 101}

\bibitem[\protect\citeauthoryear{{van Dokkum}, {Abraham}, {Merritt}, {Zhang},
  {Geha}  \& {Conroy}}{{van Dokkum} et~al.}{2015}]{vanDokkum2015}
{van Dokkum} P.~G.,  {Abraham} R.,  {Merritt} A.,  {Zhang} J.,  {Geha} M.,
  {Conroy} C.,  2015, \mn@doi [\apjl] {10.1088/2041-8205/798/2/L45}, \href
  {https://ui.adsabs.harvard.edu/abs/2015ApJ...798L..45V} {798, L45}

\makeatother
\end{thebibliography}

\label{lastpage}
\end{document}